\input harvmac

\def\hat{\widehat}

%
\let\includefigures=\iftrue
%
%
%
\newfam\black
\input rotate
\input epsf
\noblackbox
%
%
\includefigures
\message{If you do not have epsf.tex (to include figures),}
\message{change the option at the top of the tex file.}
\def\figin{\epsfcheck\figin}\def\figins{\epsfcheck\figins}
\def\epsfcheck{\ifx\epsfbox\UnDeFiNeD
\message{(NO epsf.tex, FIGURES WILL BE IGNORED)}
\gdef\figin##1{\vskip2in}\gdef\figins##1{\hskip.5in}
\else\message{(FIGURES WILL BE INCLUDED)}%
\gdef\figin##1{##1}\gdef\figins##1{##1}\fi}
\def\DefWarn#1{}
\def\N{{\cal N}}
\def\figinsert{\goodbreak\midinsert}
\def\ifig#1#2#3{\DefWarn#1\xdef#1{fig.~\the\figno}
\writedef{#1\leftbracket fig.\noexpand~\the\figno}%
\figinsert\figin{\centerline{#3}}\medskip\centerline{\vbox{\baselineskip12pt
\advance\hsize by -1truein\noindent\footnotefont{\bf
Fig.~\the\figno:} #2}}
\bigskip\endinsert\global\advance\figno by1}
\else
\def\ifig#1#2#3{\xdef#1{fig.~\the\figno}
\writedef{#1\leftbracket fig.\noexpand~\the\figno}%
\global\advance\figno by1} \fi
\def\hat{\widehat}
\def\tilde{\widetilde}

\def\yboxit#1#2{\vbox{\hrule height #1 \hbox{\vrule width #1
\vbox{#2}\vrule width #1 }\hrule height #1 }}
\def\fillbox#1{\hbox to #1{\vbox to #1{\vfil}\hfil}}
\def\ybox{{\lower 1.3pt \yboxit{0.4pt}{\fillbox{8pt}}\hskip-0.2pt}}

\def\rightarrowbox#1#2{
  \setbox1=\hbox{\kern#1{${ #2}$}\kern#1}
  \,\vbox{\offinterlineskip\hbox to\wd1{\hfil\copy1\hfil}
    \kern 3pt\hbox to\wd1{\rightarrowfill}}}

\def\p{\partial}

\def\half{{1\over 2}}

\def\tr{{\rm tr\ }}

\def\vev#1{\langle{#1}\rangle}

\def\tilde{\widetilde}

\def\II{\relax{I\kern-.10em I}}

\def\IZ{\relax\ifmmode\mathchoice
{\hbox{\cmss Z\kern-.4em Z}}{\hbox{\cmss Z\kern-.4em Z}}
{\lower.9pt\hbox{\cmsss Z\kern-.4em Z}} {\lower1.2pt\hbox{\cmsss
Z\kern-.4em Z}}\else{\cmss Z\kern-.4em Z}\fi}
\def\IB{\relax{\rm I\kern-.18em B}}
\def\IC{{\relax\hbox{$\inbar\kern-.3em{\rm C}$}}}
\def\ID{\relax{\rm I\kern-.18em D}}
\def\IE{\relax{\rm I\kern-.18em E}}
\def\IF{\relax{\rm I\kern-.18em F}}
\def\IG{\relax\hbox{$\inbar\kern-.3em{\rm G}$}}
\def\IGa{\relax\hbox{${\rm I}\kern-.18em\Gamma$}}
\def\IH{\relax{\rm I\kern-.18em H}}
\def\II{\relax{\rm I\kern-.18em I}}
\def\IK{\relax{\rm I\kern-.18em K}}
\def\IN{\relax{\rm I\kern-.18em N}}
\def\IP{\relax{\rm I\kern-.18em P}}

%
\def\inbar{\,\vrule height1.5ex width.4pt depth0pt}

\def\p{\partial}

\font\cmss=cmss10 \font\cmsss=cmss10 at 7pt
\def\IR{\relax{\rm I\kern-.18em R}}

\def\lp10{l_P^{10}}
\def\lp11{l_P^{11}}
\def\R11{R_{11}}

\def\slashed#1{#1 \!\!\! /}
\def\p{m}
\def\q{s}

\newbox\tmpbox\setbox\tmpbox\hbox{\abstractfont
}
 \Title{\vbox{\baselineskip12pt\hbox to\wd\tmpbox{\hss
 hep-th/0406177} }}
 {\vbox{\centerline{Twistor Space Structure}
 \bigskip
 \centerline{Of One-Loop Amplitudes In Gauge Theory}
 }}
\smallskip
\centerline{Freddy Cachazo$^*$, Peter Svr\v cek$^\#$, and Edward
Witten$^*$} \bigskip \centerline{\it $^*$ School of Natural
Sciences, Institute for Advanced Study, Princeton NJ 08540
USA}\bigskip \centerline{\it $^\#$ Department of Physics, Joseph
Henry Laboratories, Princeton NJ 08540 USA}
\bigskip
\vskip 1cm \noindent

\input amssym.tex
We analyze the twistor space structure of certain one-loop
amplitudes in gauge theory.  For some amplitudes, we find
decompositions that make the twistor structure manifest; for
others, we explore the twistor space structure by finding
differential equations that the amplitudes obey.

\Date{June 2004}

\lref\BernZX{ Z.~Bern, L.~J.~Dixon, D.~C.~Dunbar and
D.~A.~Kosower, ``One Loop N Point Gauge Theory Amplitudes,
Unitarity And Collinear Limits,'' Nucl.\ Phys.\ B {\bf 425}, 217
(1994), hep-ph/9403226.
}

\lref\BernCG{ Z.~Bern, L.~J.~Dixon, D.~C.~Dunbar and
D.~A.~Kosower, ``Fusing Gauge Theory Tree Amplitudes into Loop
Amplitudes,'' Nucl.\ Phys.\ B {\bf 435}, 59 (1995),
hep-ph/9409265.
}

\lref\WittenNN{ E.~Witten, ``Perturbative Gauge Theory as a String
Theory in Twistor Space,'' hep-th/0312171.
}

\lref\CachazoKJ{ F.~Cachazo, P.~Svrcek and E.~Witten, ``MHV
Vertices and Tree Amplitudes in Gauge Theory,'' hep-th/0403047.
}

\lref\berkwitten{N. Berkovits and E. Witten,  ``Conformal
Supergravity In Twistor-String Theory,'' hep-th/0406051.}

\lref\penrose{R. Penrose, ``Twistor Algebra,'' J. Math. Phys. {\bf
8} (1967) 345.}

\lref\BernSX{ Z.~Bern, L.~J.~Dixon and D.~A.~Kosower, ``New QCD
Results From String Theory,'' hep-th/9311026.
}

\lref\berends{F. A. Berends, W. T. Giele and H. Kuijf, ``On
Relations Between Multi-Gluon And Multi-Graviton Scattering,"
Phys. Lett {\bf B211} (1988) 91.}

\lref\berendsgluon{F. A. Gerends, W. T. Giele and H. Kuijf,
``Exact and Approximate Expressions for Multigluon Scattering,"
Nucl. Phys. {\bf B333} (1990) 120.}

\lref\bernplusa{Z. Bern, L. Dixon and D. A. Kosower, ``New QCD
Results From String Theory,'' in {\it Strings '93}, ed. M. B.
Halpern et. al. (World-Scientific, 1995), hep-th/9311026.}

\lref\bernplusb{Z. Bern, G. Chalmers, L. J. Dixon and D. A.
Kosower, ``One Loop $N$ Gluon Amplitudes with Maximal Helicity
Violation via Collinear Limits," Phys. Rev. Lett. {\bf 72} (1994)
2134.}

\lref\bernfive{Z. Bern, L. J. Dixon and D. A. Kosower, ``One Loop
Corrections to Five Gluon Amplitudes," Phys. Rev. Lett {\bf 70}
(1993) 2677.}

\lref\bernfourqcd{Z.Bern and  D. A. Kosower, "The Computation of
Loop Amplitudes in Gauge Theories," Nucl. Phys.  {\bf B379,}
(1992) 451.}

\lref\cremmerlag{E. Cremmer and B. Julia, ``The $N=8$ Supergravity
Theory. I. The Lagrangian," Phys. Lett.  {\bf B80} (1980) 48.}

\lref\cremmerso{E. Cremmer and B. Julia, ``The $SO(8)$
Supergravity," Nucl. Phys.  {\bf B159} (1979) 141.}

\lref\dewitt{B. DeWitt, "Quantum Theory of Gravity, III:
Applications of Covariant Theory," Phys. Rev. {\bf 162} (1967)
1239.}

\lref\dunbarn{D. C. Dunbar and P. S. Norridge, "Calculation of
Graviton Scattering Amplitudes Using String Based Methods," Nucl.
Phys. B {\bf 433,} 181 (1995), hep-th/9408014.}

\lref\ellissexton{R. K. Ellis and J. C. Sexton, "QCD Radiative
corrections to parton-parton scattering," Nucl. Phys.  {\bf B269}
(1986) 445.}

\lref\gravityloops{Z. Bern, L. Dixon, M. Perelstein, and J. S.
Rozowsky, ``Multi-Leg One-Loop Gravity Amplitudes from Gauge
Theory,"  hep-th/9811140.}

\lref\kunsztqcd{Z. Kunszt, A. Singer and Z. Tr\'{o}cs\'{a}nyi,
``One-loop Helicity Amplitudes For All $2\rightarrow2$ Processes
in QCD and ${\cal N}=1$ Supersymmetric Yang-Mills Theory,'' Nucl.
Phys.  {\bf B411} (1994) 397, hep-th/9305239.}

\lref\mahlona{G. Mahlon, ``One Loop Multi-photon Helicity
Amplitudes,'' Phys. Rev.  {\bf D49} (1994) 2197, hep-th/9311213.}

\lref\mahlonb{G. Mahlon, ``Multi-gluon Helicity Amplitudes
Involving a Quark Loop,''  Phys. Rev.  {\bf D49} (1994) 4438,
hep-th/9312276.}

\lref\klt{H. Kawai, D. C. Lewellen and S.-H. H. Tye, ``A Relation
Between Tree Amplitudes of Closed and Open Strings," Nucl. Phys.
{B269} (1986) 1.}

\lref\pppmgr{Z. Bern, D. C. Dunbar and T. Shimada, ``String Based
Methods In Perturbative Gravity," Phys. Lett.  {\bf B312} (1993)
277, hep-th/9307001.}

\lref\GiombiIX{ S.~Giombi, R.~Ricci, D.~Robles-Llana and
D.~Trancanelli, ``A Note on Twistor Gravity Amplitudes,''
hep-th/0405086.
}

\lref\WuFB{ J.~B.~Wu and C.~J.~Zhu, ``MHV Vertices and Scattering
Amplitudes in Gauge Theory,'' hep-th/0406085.
}

\lref\Feynman{R.P. Feynman, Acta Phys. Pol. 24 (1963) 697, and in
{\it Magic Without Magic}, ed. J. R. Klauder (Freeman, New York,
1972), p. 355.}

\lref\Peskin{M.E. Peskin and D.V. Schroeder, {\it An Introduction
to Quantum Field Theory} (Addison-Wesley Pub. Co., 1995).}

\lref\parke{S. Parke and T. Taylor, ``An Amplitude For $N$ Gluon
Scattering,'' Phys. Rev. Lett. {\bf 56} (1986) 2459; F. A. Berends
and W. T. Giele, ``Recursive Calculations For Processes With $N$
Gluons,'' Nucl. Phys. {\bf B306} (1988) 759. }


\newsec{Introduction}

Perturbative gauge theory has many remarkable properties, among
them the surprising accessability and elegant structure of certain
one-loop $S$-matrix elements \refs{\BernZX,\BernCG}.  Some of the
unexpected simplicity of perturbative gauge theory can be
explained by reinterpreting this subject in terms of a topological
string theory with twistor space as the target.  This was proposed
in \WittenNN, where more detailed references concerning
perturbative gauge theory and twistor space can be found.

Even if one does not know a twistor-string theory appropriate for
computing a given scattering amplitude, or does not understand it
properly, one can, as explained in section 3 of \WittenNN, gain
some insight  by studying the differential equations that the
scattering amplitudes obey. With this in mind, we have undertaken
a detailed study of differential equations obeyed by tree level
and one-loop scattering amplitudes in gauge theories with varying
degrees of supersymmetry.  The tree level results suggested an
interpretation via ``MHV tree diagrams'' that we have presented
separately \CachazoKJ.  The present paper aims to explain the
one-loop results.

For gluon scattering at tree level (in renormalizable gauge
theories), supersymmetry does not matter.  For loop amplitudes, it
does.  Not all one-loop amplitudes have been computed. Roughly
speaking, in this paper we consider nearly all of the available
one-loop gluon scattering amplitudes with massless internal lines.
We study MHV (maximally helicity violating) one-loop amplitudes in
theories with ${\cal N}=4$ or ${\cal N}=1$ supersymmetry
\refs{\BernZX,\BernCG}, and certain nonsupersymmetric one-loop
amplitudes \BernCG.

\ifig\tomu{Shown here are twistor configurations that we find
contribute to one-loop supersymmetric MHV amplitudes.  In (a), all
gluons are inserted on a pair of disjoint lines.  In (b), all
gluons are inserted on a pair of intersecting lines.  (c) is just
like (b) except that one gluon is inserted not on the pair of
intersecting lines but somewhere else in the plane containing the
two lines. In the figures, dashed lines indicate twistor space
propagators whose presence we conjecture, though it is not
directly revealed by calculations in this paper. }
{\epsfxsize=0.90\hsize\epsfbox{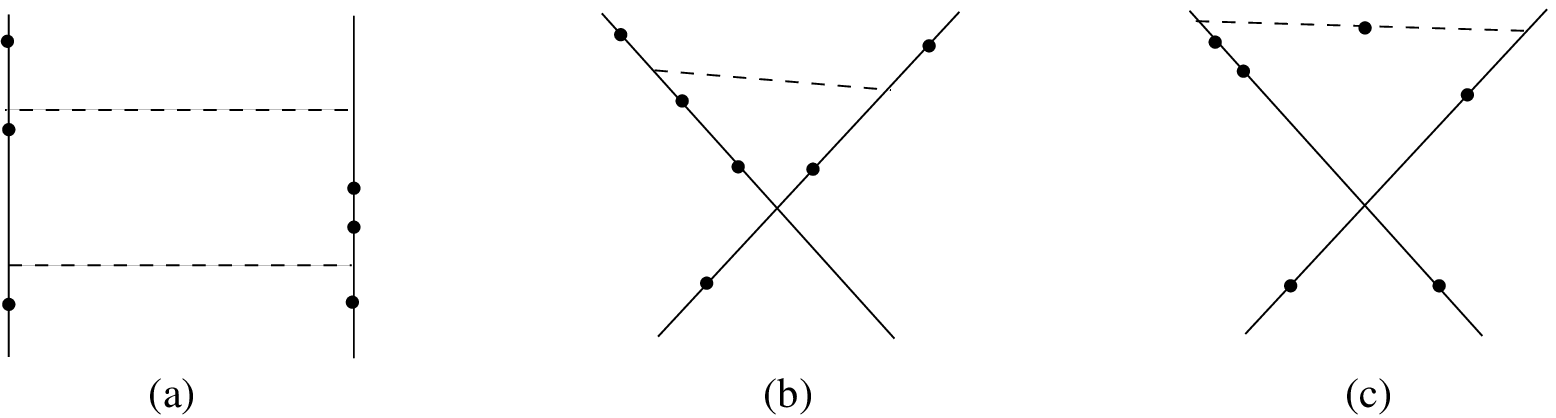}}

Our results for the supersymmetric one-loop MHV amplitudes  are
qualitatively similar to what one would guess based on the
twistor-string conjecture in \WittenNN, but there is an apparent
discrepancy, whose meaning will be clarified in a separate paper.
From \WittenNN, one would anticipate two possible types of twistor
space contribution to a one-loop $n$-gluon MHV scattering process.
In one configuration, sketched in \tomu(a), all $n$ gluons are
supported on a pair of lines in twistor space; the lines are
connected by two twistor propagators. In the second configuration,
sketched in \tomu(b), all $n$ gluons are supported on a curve $C$
of genus zero and degree two; there is also a twistor space
propagator connecting this curve to itself. Our study of the
differential equations, however, has revealed that $C$ reduces to
a pair of intersecting lines, and that the conditions in \tomu(b)
need to be somewhat relaxed. The supersymmetric one-loop MHV
scattering amplitudes actually appear to receive contributions of
the type indicated in \tomu(c), with only $n-1$ of the gluons
contained in the two lines. The two intersecting lines are
automatically contained in a $\Bbb{CP}^2\subset\Bbb{CP}^3$, and
the $n^{th}$ gluon is contained in this $\Bbb{CP}^2$ (with
``derivative of a delta function'' support).

We do not know what kind of twistor-string theory would generate
this structure, but we hope that our result may serve as a useful
clue.  There may be an important difference between pure super
Yang-Mills theory (which we study here), and super Yang-Mills
theory coupled to conformal supergravity, which  \berkwitten\ is
described by currently known forms of twistor-string theory.

In section 2, we briefly review the use of differential equations
to investigate the twistor space structure of scattering
amplitudes, and explain how analysis of these equations served as
a clue to the description of Yang-Mills tree amplitudes via MHV
tree diagrams \CachazoKJ. In section 3, we study the one-loop MHV
amplitudes in ${\cal N}=4$ super Yang-Mills theory. Our main
result is a noncovariant decomposition of the amplitude that makes
almost manifest the differential equations obeyed by the amplitude
and its twistor space structure. The decomposition is somewhat
similar to the one we found for tree amplitudes in our previous
paper, but in contrast to that case, we are unfortunately not able
to give a simple explanation of what the pieces mean. In section
4, we study the one-loop MHV amplitudes in ${\cal N}=1$ super
Yang-Mills theory.  We find that, at least for one-loop MHV
amplitudes, the twistor space structure for ${\cal N}=1$  seems to
be nearly the same as for ${\cal N}=4$. In section 5, we study
some nonsupersymmetric one-loop amplitudes. Again, analysis of
differential equations suggests a twistor space structure that is
surprisingly similar to the supersymmetric case. The most
important difference may be that in nonsupersymmetric gauge
theories, there is a one-loop  amplitude for gluons all of
positive helicity that must be included as a new building block
alongside the MHV tree amplitudes.

As we will explain in a separate paper, the configuration of
\tomu(c) arises, in a certain sense, from a holomorphic anomaly in
evaluating the amplitude derived from a twistor space
configuration of type 1(a). When this is taken into account, one
can possibly salvage the naive twistor space viewpoint of figures
1(a) and 1(b).

\newsec{Review Of Differential Equations And Tree Level
Amplitudes}

Scattering amplitudes are usually described for particles of
definite momentum $p_\mu$.  For a massless particle, one can
factor the momentum in terms of spinors; setting $p_{a\dot
a}=\sigma_{a\dot a}^\mu p_\mu$, one has $p_{a\dot
a}=\lambda_a\tilde\lambda_{\dot a}$, with $\lambda_a$ and
$\tilde\lambda_{\dot a}$ being spinors of, respectively, positive
and negative helicity.  Spinor inner products are defined by
$\langle
\lambda,\lambda'\rangle=\epsilon_{ab}\lambda^a\lambda'{}^b$,
$[\tilde\lambda,\tilde\lambda']=\epsilon_{\dot a\dot
b}\tilde\lambda^{\dot a}\tilde\lambda'{}^{\dot b}$. We also write
$\langle i,j\rangle$ for $\langle\lambda_i,\lambda_j\rangle$, and
$[i,j]$ for $[\tilde\lambda_i,\tilde\lambda_j]$. For further
details of spinor notation, see section 2 of \WittenNN.

Twistor space \penrose\ is introduced by a Fourier transform from
$\tilde\lambda^{\dot a}$ to a new variable  $\mu^{\dot a}$.  (See
\WittenNN\ for a detailed description of this Fourier transform.)
One interprets $Z^I=(\lambda^a,\mu^{\dot a})$ as homogeneous
coordinates of a complex projective space $\Bbb{CP}^3$ that is
known as twistor space. (Twistor space also has a supersymmetric
extension, but this will not be needed here.)

In a twistor description, instead of describing the $i^{th}$
external massless particle in an $n$-particle scattering amplitude
by its momentum $p_i^{a\dot a}=\lambda_i^a\tilde\lambda_i^{\dot
a}$, one associates it with a point $Z_i$ in twistor space with
homogeneous coordinates $Z_i^I=(\lambda_i^a,\mu_i^{\dot a})$. The
twistor space scattering amplitude is a function (or $(0,n)$-form)
$\tilde A(Z_1^I,\dots,Z_n^I)$ on  $(\Bbb{CP}^3)^n$, one copy of
$\Bbb{CP}^3$ for each external particle.  By ``determining the
twistor space structure'' of the scattering amplitude, we mean
determining the smallest algebraic subspace of $(\Bbb{CP}^3)^n$ on
which $\tilde A(Z_1^I,\dots,Z_n^I)$ is supported.  The answer is
interesting if this subspace has a small dimension and, hopefully,
a simple string theory interpretation.

\subsec{Building Blocks Of Differential Equations}

A variety of conditions on collections of points in $\Bbb{CP}^3$
were considered in section 3 of \WittenNN.  Happily, in the
present paper we need only the simplest of those conditions. Given
three points  $P_i,P_j,P_k\in \Bbb{CP}^3$ with homogeneous
coordinates $Z_i^I$, $Z_j^I$, and $Z_k^I$, the condition that they
lie on a ``line,'' that is a linearly embedded copy of
$\Bbb{CP}^1$, is that $F_{ijkL}=0$, where
\eqn\murov{F_{ijkL}=\epsilon_{IJKL}Z_i^IZ_j^JZ_k^K.} And given
four points $P_i,P_j,P_k,P_l\in \Bbb{CP}^3$, the condition that
they are all contained in a ``plane,'' that is a linearly embedded
copy of $\Bbb{CP}^2\subset \Bbb{CP}^3$, is that $K_{ijkl}=0$,
where \eqn\urov{K_{ijkl}=\epsilon_{IJKL}Z_i^IZ_j^JZ_k^KZ_l^L.}

Suppose that we are presented with the momentum space version of a
scattering amplitude.  We describe it in terms of spinors as a
function $A(\lambda_1^a,\tilde\lambda_1^{\dot
a};\dots;\lambda_n^a,\tilde\lambda_n^{\dot a})$.  The condition
that the equivalent twistor space amplitude has support where the
points $i,j,k$ are collinear is that $F_{ijkL}A=0$.  Here
$F_{ijkL}$ is interpreted as a differential operator (acting on a
function of $\lambda,\tilde\lambda$ rather than $\lambda,\mu$) via
$\mu\to i\partial/\partial\lambda$.  It is often useful to
abbreviate the statement that $F_{ijkL}A=0$ for all $L$ and write
simply $F_{ijk}A=0$.  Likewise, the condition that the twistor
space amplitude has support where the points $i,j,k,l$ are
coplanar is that $K_{ijkl}A=0$, where $K_{ijkl}$ is similarly
interpreted as a differential operator. This process of
interpreting a function of twistor coordinates as a differential
operator on momentum variables is implemented in many examples in
section 3 of \WittenNN.

We can give a few simple criteria for an amplitude to be
annihilated by the collinear operator $F_{ijkL}$.  Setting $L=\dot
a$, the operator  $F_{ijk \dot a}$ is concretely \eqn\nnin{\langle
i,j\rangle {\partial\over
\partial\tilde\lambda_k^{\dot a}}+\langle
j,k\rangle{\partial\over\partial\tilde\lambda_i^{\dot a}} +\langle
k,i\rangle {\partial\over\partial\tilde\lambda_j^{\dot a}}.}
Obviously, this operator annihilates any scattering amplitude $A$
 that depends on the $\lambda$'s but is independent of
the $\tilde\lambda$'s.  This is likewise true of the $L=a$
components of $F_{ijkL}$, which are homogeneous and quadratic in
$\partial/\partial\tilde\lambda$. So if the scattering amplitude
is independent of $\tilde\lambda_i$, $\tilde\lambda_j$, and
$\tilde\lambda_k$, then particles $i,j,k$ are supported on a line
in twistor space.

An amplitude that depends only on the $\lambda$'s and not the
$\tilde\lambda$'s is often said to be ``holomorphic.'' The
motivation for this terminology comes from considering physical
scattering amplitudes in Minkowski spacetime. For real momenta in
Lorentz signature, $\tilde\lambda$ is the complex conjugate of
$\lambda$ (up to a sign that depends on the sign of the energy),
so in real Minkowski spacetime, an amplitude is holomorphic in
$\lambda$ precisely if it is independent of $\tilde\lambda$.

More generally, if the scattering amplitude
$A(\lambda,\tilde\lambda)$ is polynomial in  $\tilde\lambda_i$,
$\tilde\lambda_j$, and $\tilde\lambda_k$, then it is annihilated
by $F_{ijk}^s$ (that is, by all components of
$F_{ijkL_1}F_{ijkL_2}\dots F_{ijkL_s}$) for some integer $s$. In
this situation, as explained in \WittenNN, particles $i,j$, and
$k$ are still supported on a line in twistor space, but now with
``derivative of a delta function'' (or multiple derivative of  a
delta function) support. A recent computation \berkwitten\ of tree
level MHV scattering involving supergravitons gives one example of
how such polynomial dependence on $\tilde\lambda$'s can arise; in
that example, the factors of $\tilde\lambda$ come from the
structure of the vertex operators.

All this has another important generalization.  Let $P^{a\dot
a}=p_i^{a\dot a}+p_j^{a\dot a}+p_k^{a\dot a}$ be the sum of the
momenta of particles $i,j,$ and $k$.  Then $F_{ijk}$ annihilates
any amplitude $A(\lambda_i,\lambda_j,\lambda_k;P)$ that depends on
$\tilde\lambda_i$, $\tilde\lambda_j$, and $\tilde\lambda_k$ only
via $P$.  For example, to verify that $F_{ijk\dot a}A=0$, after
using the chain rule, we need \eqn\tomogo{\langle
\lambda_i,\lambda_j\rangle\lambda_k^a+\langle\lambda_j,\lambda_k\rangle
\lambda_i^a+\langle\lambda_k,\lambda_i\rangle\lambda_j^a=0.} This
identity holds because the quantity on the left takes values in a
two-dimensional vector space, and is trilinear and antisymmetric
in the three vectors $\lambda_i$, $\lambda_j$, and $\lambda_k$.

\subsec{Examples}

Now let us discuss some examples.  In all cases, we consider
single-trace amplitudes with $n$ gluons in cyclic order $123\dots
n$. We begin with the five gluon tree amplitude
$A(\lambda_1,\tilde\lambda_1;\dots;\lambda_5,\tilde\lambda_5)$
with three gluons of negative helicity.
 Differential equations
obeyed by this amplitude were described in section 3 of \WittenNN.

There are two types of equation.  One asserts that the amplitude
is supported on configurations for which three consecutive points
out of the five are contained in a ``line,'' that is a
$\Bbb{CP}^1$.  The other two points are automatically on a line
(as there is a straight line through any two points), so the five
points are actually on a union of two lines.  The system of
differential equations which exhibits this fact is
\eqn\nno{\prod_{k=1}^5W_{kI_k}A=0,} where $W_{kI}=F_{k-1,k,k+1I}$
annihilates amplitudes for which particles $k-1,k,$ and $k+1$ are
collinear. Eqn. \nno\ holds for arbitrary choices of the indices
$I_k$, and we abbreviate it by writing $\prod_{k=1}^5W_kA=0$.

In twistor space, where the $W_k$ are simply multiplication
operators, the assertion that $A$ is annihilated by
$\prod_{k=1}^5W_k$ means that it is supported on the subset of
$(\Bbb{CP}^3)^5$ on which at least one of the $W_k$ vanishes, that
is, on which at least three consecutive gluons are collinear.

The other differential equation obeyed by $A$ was found to be that
$K_{ijkl}A=0$ for all $i,j,k$, and $l$.  This asserts that the
amplitude is supported on configurations of five points that lie
in a common $\Bbb{CP}^2$.  (The equation $K_{ijkl}A=0$ is actually
only valid for generic momenta; delta function terms  enter this
equation, as was anticipated to some extent in section 3 of
\WittenNN\ and as we explain below.)

\ifig\yoye{(a) A pair of intersecting lines. (b) The quiver
corresponding to (a).  Each vertex in the quiver represents a
line; two vertices are connected if and only if the lines
intersect.} {\epsfxsize=0.55\hsize\epsfbox{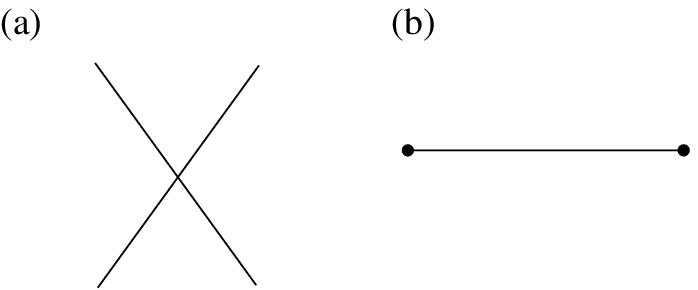}}

Two lines in three-space intersect if and only if they are
contained in a common plane, which in the present context means a
common $\Bbb{CP}^2$.  So the two sets of equations, taken
together, mean that the five gluons are inserted on a pair of
intersecting lines. The relevant configurations are indicated in
\yoye(a).

We associate the configuration of \yoye(a) with a certain
``quiver.'' A quiver is just a graph containing points or
vertices, some of which are connected by lines. We restrict
ourselves to connected quivers, and for quivers related to tree
diagrams, we want graphs that contain no closed loops.  To make a
quiver from a configuration of lines in twistor space, we draw a
vertex for every line, and we connect two vertices if and only if
the corresponding lines intersect.  Thus, the configuration in
\yoye(a) corresponds to the simple quiver in \yoye(b).

\ifig\qime{(a) The two tree-level quivers with four vertices. (b)
An arrangement of gluons corresponding to the first quiver in (a).
 Each vertex in the quiver is represented by a disc, a line joining vertices is
 represented by a thin strip connecting two discs, and the gluons are arranged on
 the boundaries of the discs.  Shown is an arrangement contributing to a single-trace
 amplitude with eight gluons.} {\epsfxsize=0.90\hsize\epsfbox{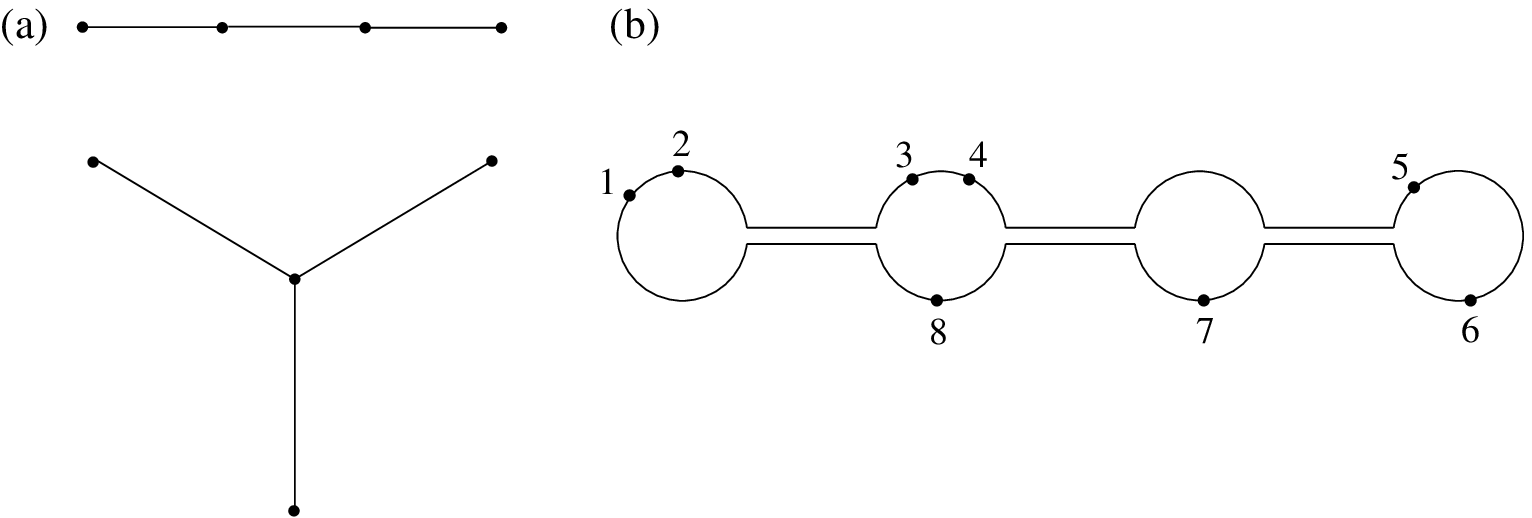}}

Now we can describe the results of our study of differential
equations obeyed by  Yang-Mills tree amplitudes with many gluons:
the tree amplitudes in general are supported on quivers in twistor
space. In the case of an $n$-gluon amplitude with $q$ gluons of
negative helicity, the quiver is constructed from $d=q-1$
intersecting lines. In general, all possible topologies for the
quivers must be included. (The first case with more than one
possible quiver is $d=4$, where there are two possibilities, as
indicated in \qime(a).)  For each quiver, one must sum over
different arrangements of particles among the various lines (or
$\Bbb{CP}^1$'s) in the quiver. The allowed arrangements can be
motivated by a hypothetical duality between twistor-string theory
and a Type I string theory based on Chan-Paton factors.  We
imagine replacing each $\Bbb{CP}^1$ by a small disc, and each
intersection of $\Bbb{CP}^1$'s by a thin strip connecting the
discs. Finally, we sum over all arrangements of external gluons on
the boundaries of the discs that are compatible with the cyclic
ordering $123\dots n$.  For example, for the case that the number
of gluons is $n=8$, one possible arrangement corresponding to the
first quiver in \qime(a) is shown in \qime(b).

To show that the amplitude is supported on the quiver, we do the
following.  Let $T$ be the set of  possible arrangements on
possible quivers, with the quivers and arrangements constructed by
the rules of the last paragraph. For each $t\in T$, pick an
operator ${\cal D}_t$ that should annihilate a configuration in
twistor space associated with that arrangement. For example, if
$t$ corresponds to the arrangement of eight gluons in \qime(b), we
could take ${\cal D}_t$ to be $F_{348}$ or $K_{1234}$ (since in
this configuration, particles $3,4,$ and 8 are contained in a
line, and particles 1,2,3, and 4 are contained in a pair of
intersecting lines and hence in a plane).  Then the claim is that
\eqn\nonno{\prod_{t\in T}{\cal D}_tA=0.} This should hold for all
choices of ${\cal D}_t$.

Even for modest values of $n$ and $q$, the differential equations
in \nonno\ are very numerous (as there are many choices of the
${\cal D}_t$) and of very high order. Nevertheless, for a certain
range of $n$ and $q$, sufficient to be convincing, we established
with some computer assistance that Yang-Mills tree amplitudes obey
\nonno\ and are annihilated by no other differential operators
that can be expressed as products of a comparably small number of
$F$'s and $K$'s. The quiver picture is more restrictive than what
we originally anticipated based on \WittenNN, and was discovered
by trial and error with the differential equations.

Happily, we need not explain any further the details of these tree
level differential equations, because there is a more transparent
way to understand the quiver picture: it motivated the concept of
MHV tree diagrams \CachazoKJ, which in fact make the quiver
picture manifest.

\ifig\fori{Two ``lines,'' that is two $\Bbb{CP}^1$'s, with gluon
insertions on them, connected by a twistor space propagator and
contributing to a tree-level five-gluon amplitude.}
{\epsfxsize=0.75\hsize\epsfbox{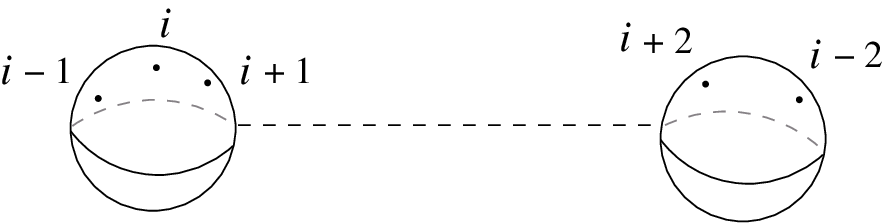}}

To illustrate the point, we reconsider the five-gluon amplitude
with three gluons of negative helicity. The extension of this
discussion to general MHV tree diagrams and the associated quivers
will hopefully be clear. In the approach via MHV tree diagrams,
the five-gluon amplitude is obtained as the sum of five MHV tree
diagrams, one of which is indicated in \fori.  Each diagram
contains a pair of MHV vertices connected by a propagator. One
vertex contains (say) gluons $i-1, $ $i $, and $i+1$ and the other
contains the other two gluons.   The two MHV vertices form the
vertices of a quiver (which is simply the quiver of fig. 2(a)),
and the choice of how to assign particles to the different MHV
vertices  corresponds precisely to the choice of arrangement of
particles on this quiver.

The propagator in the MHV tree diagram is $1/P^2$, where
$P=p_{i-1}+p_i+p_{i+1}$ is the total momentum flowing between the
vertices.  We claim that the amplitude of this particular MHV tree
diagram is supported on configurations in which gluons $i-1$, $i$,
and $i+1$ are collinear. Indeed, the amplitude is annihilated by
$W_i$ since the criterion of section 2.1 is satisfied:  each MHV
vertex depends on the $\tilde\lambda$'s only through $P$, while
the propagator depends only on $P^2$. The five-gluon amplitude is
a sum of five MHV tree diagrams each of which is annihilated by
one of the $W_i$, so the total amplitude is annihilated by the
product $W_1W_2W_3W_4W_5$, a statement that is part of the quiver
picture.

The remainder of the quiver picture for the five-gluon amplitude
is the assertion that this amplitude is annihilated by the
operators $K_{ijkl}$ that measure coplanarity of four points in
twistor space.  Each MHV diagram separately has this property; it
is instructive to verify this directly by writing $K$ as a
differential operator that acts on the amplitude derived from an
MHV tree diagram. But another approach to explaining the statement
is more illuminating, and also shows the limitations of the quiver
picture.

To get the physical scattering amplitudes, the propagator in an
MHV tree diagram should be  $i/(P^2+i\epsilon)$. In coordinate
space, the Fourier transform of this function is the Feynman
propagator $D_F(x)$, which has non-zero support inside and outside
the light cone (for example, see section 2.4 of \Peskin).
However, with a different $i\epsilon$ prescription, the Fourier
transform of $1/P^2$ in four dimensions has its support on the
light cone.

To explain this, we start with a massless scalar field $\phi$. The
retarded propagator is defined as $D_R(x)
=\vartheta(x^0)\langle\Omega |[\phi(x),\phi(0)]|\Omega\rangle$,
where $|\Omega\rangle $ is the vacuum state, and $\vartheta(x^0)$
is equal to 1 for $x^0\geq 0$ and vanishes for $x^0<0$.  The
retarded propagator in $n$ dimensions obeys \eqn\mucno{\square
D_R(x)=-i\delta^n(x),} where $\square$ is the massless wave
operator. In any dimension and for a particle of any mass,
$D_R(x)$ vanishes outside the light cone, by virtue of causality,
and for $x^0<0$, because of the factor of $\vartheta(x^0)$ in its
definition.  For a massless particle in an even spacetime
dimension, $D_R(x)$ also vanishes {\it inside} the light cone; it
is supported entirely on the future light cone. (In four
dimensions, this statement is an aspect of Huygen's principle; the
light signal observed at a given point in spacetime depends only
on the sources {\it on} the past light cone, not on sources {\it
inside} the past light cone.)

By virtue of \mucno, the Fourier transform of $D_R(x)$ is $i/P^2$,
with some way of treating the singularity at $P^2=0$. Therefore,
if we use the retarded propagator\foot{This and subsequent
statements also hold if we use the advanced propagator or a
half-retarded, half-advanced propagator.} in an MHV tree diagram,
rather than the Feynman propagator, the propagation will occur
only on the light cone.  By contrast, the Feynman propagator
$D_F(x)$, whose Fourier transform is $i/(P^2+i\epsilon)$,
describes propagation inside, outside, or on the light cone.

The vertices in the MHV tree diagram of figure 4 represent lines
(or $\Bbb{CP}^1$'s) in twistor space that correspond to points in
Minkowski space.  From what we have just said, if we use the
retarded propagator, these points are at lightlike separation in
Minkowski space.  In the twistor transform, points in Minkowski
space that are at lightlike separation correspond to
$\Bbb{CP}^1$'s that intersect.\foot{Let one $\Bbb{CP}^1$ be given
by  $\mu^{\dot a}=x^{a\dot a}\lambda_a$ and the other by
$\mu^{\dot a}=y^{a\dot a}\lambda_a$.  The difference $y-x$ is
lightlike if and only if $\det\,(y_{a\dot a}-x_{a\dot a})=0$,
which is the condition for existence of $\lambda$ such that
$(y^{a\dot a}-x^{a\dot a})\lambda_a=0$.  Precisely when that is
so, the two $\Bbb{CP}^1$'s intersect at that value of $\lambda$
and $\mu^{\dot a}=x^{a\dot a}\lambda_a=y^{a\dot a}\lambda_a$.} So
if we use the retarded propagator, the $\Bbb{CP}^1$'s
corresponding to the vertices intersect and are thus coplanar.

Since the correct physical amplitude is derived from the Feynman
propagator rather than the retarded propagator, one should
hesitate to claim that the lines intersect and the amplitudes are
supported on a quiver. The two propagators differ in momentum
space by delta function terms supported at $P^2=0$ -- terms which
we usually overlook in studying the differential equations.  Our
argument shows that if (and only if) one ignores such delta
function terms
 in evaluating $K_{ijkl}A$, one will get $K_{ijkl}A=0$.

In \Feynman, Feynman decomposes one-loop amplitudes by expressing
the Feynman propagator as the retarded propagator plus a function
that is supported on-shell.  We believe that this decomposition
may be related to the twistor space decomposition displayed in
figure 1.

\bigskip\noindent{\it Conclusion}

We have explained why and to what extent the quiver picture is
true. Despite its limitations, the quiver picture was an important
clue to the understanding of MHV tree diagrams that we have
presented elsewhere \CachazoKJ.  In the rest of this paper, we
attempt to determine the one-loop analog of the quiver picture (at
least for some classes of one-loop amplitudes), hoping that this
will similarly serve as a useful clue to a better understanding of
twistor-string theory.

\newsec{$\N =4$ One-loop MHV Amplitudes}

\subsec{Description Of The Amplitudes}

One-loop amplitudes with external gluons are notoriously difficult
to compute by evaluating Feynman diagrams directly. Often a
simpler problem is to find the discontinuities along the branch
cuts of the amplitude, in other words, the unitarity cuts. In
general, knowing the cuts does not fix the amplitude, for there
can be single-valued functions which lack cuts. However, Bern,
Dixon, Dunbar, and Kosower \BernZX\ have shown that some
amplitudes in gauge theories can be completely determined by their
four-dimensional cuts; these are called ``cut-constructible"
amplitudes.

Remarkably, all ${\cal N}=4$ one-loop amplitudes turn out to be
``cut-constructible" \BernZX. This result made it possible to
compute, at one-loop order, MHV amplitudes with any number of
gluons \BernZX, and also  the six-gluon  amplitudes with any
helicities \BernCG.  In this section, we analyze the twistor space
structure of the ${\cal N}=4$ one-loop MHV amplitudes. More
precisely, we consider the leading-color partial amplitudes, i.e.,
the single-trace contributions. However, it turns out that
multi-trace partial amplitudes are given as combinations of the
leading-color partial amplitudes (see section 7 of \BernZX).
Therefore our conclusions about the leading-color partial
amplitudes are also valid for the total amplitude.

MHV one-loop amplitudes (unlike more general one-loop amplitudes
in the ${\cal N}=4$ theory) have a simple dependence on the
helicity of the external gluons.  In fact, one-loop MHV $n$-gluon
amplitudes $A_n^{\rm 1-loop}$  can be written
\eqn\heli{ A_n^{\rm 1-loop} = A_n^{\rm tree} V_n.}
 $A_n^{\rm tree}$ is the familiar Parke-Taylor tree level MHV
amplitude \parke, which contains the information about the
helicities, while $V_n$ is a universal one-loop function
independent of which two gluons have negative helicity.\foot{In
our formulas, we will omit a numerical factor, usually called
$c_\Gamma$, that depends only on the dimensional regularization
parameter $\epsilon = (4-D)/2$ and is not relevant to our
analysis.}

Because $A_n^{\rm tree}$ is a holomorphic function of positive
chirality spinor variables $\lambda_i$, $i=1,\ldots ,n$, it will
really not affect our analysis. The key point is to study the
one-loop function $V_n$.

The universality of $V_n$ implies that it is invariant under
cyclic permutations of the gluons. The evaluation of $V_n$
\BernZX\ shows that it can be naturally written as a sum of terms
in which $r$ consecutive gluons, say $i,i+1,\dots,i+r-1$ (for some
$i$ and $r$), combine together in a certain sense, as do $n-2-r$
other gluons, which in this case are $i+r+1,\dots,i-2$.  Two
gluons, labeled $i-1$ and $i+r$, remain uncombined. (For a
pictorial representation, see Appendix A.) The resulting
contribution is symmetric under exchange of the two sets of
``combined'' gluons, so we can restrict to $r\leq (n-2)/2$. $V_n$
is written as a sum over such choices of $i$ and $r$:
\eqn\vampl{ V_n = \sum_{i=1}^n \sum_{r=1}^{\left[{n\over 2}\right]
-1}\left( 1 - \half\delta_{n/2 -1 , r}\right) F_{n:r;i}^{\rm
2m~e}.}
Here $F_{n:r;i}^{\rm 2m~e}$ is known as the box function and it is
given by
\eqn\scalar{\eqalign{  F_{n:r;i}^{\rm 2m~e} & = -{1\over
\epsilon^2}\left[ (-t_{i-1}^{[r+1]})^{-\epsilon} +
(-t_{i}^{[r+1]})^{-\epsilon} -(-t_{i}^{[r]})^{-\epsilon}
-(-t_{i+r+1}^{[n-r-2]})^{-\epsilon}\right] \cr & + {\rm Li}_2
\left( {1-{t_i^{[r]}\over t_{i-1}^{[r+1]}}} \right) + {\rm Li}_2
\left( {1-{t_i^{[r]}\over t_{i}^{[r+1]}}} \right)  \cr & + {\rm
Li}_2 \left( {1-{t_{i+r+1}^{[n-r-2]}\over t_{i-1}^{[r+1]}}}
\right) + {\rm Li}_2 \left( {1-{t_{i+r+1}^{[n-r-2]}\over
t_{i}^{[r+1]}}} \right) \cr & - {\rm Li}_2 \left(
{1-{t_{i}^{[r]}t_{i+r+1}^{[n-r-2]}\over
t_{i-1}^{[r+1]}t_{i}^{[r+1]}}} \right) + {1\over 2}\log^2\left(
{t_{i-1}^{[r+1]}\over t^{[r+1]}_{i}} \right).}  } Here we define
$t_i^{[k]}=(p_i+p_{i+1}+\dots +p_{i+k-1})^2$.  An explanation of
why this function is called the box function, as well as the
reason for the superscript $({\rm 2m~e})$, can be found in
Appendix A.

In this formula, ${\rm Li}_2(x)$ denotes the dilogarithm function
as defined by Euler, i.e., ${\rm Li}_2(x)=-\int_0^x \log(1-z)\;
dz/z$. $F_{n:r;i}^{\rm 2m~e}$ comes from a divergent integral (see
appendix) and $\epsilon = (4-D)/2$ is the dimensional
regularization parameter. For $k=1$, the meaning of
$(-t_{i}^{[k]})^{-\epsilon}$ needs to be clarified; one defines
$(-t_{i}^{[1]})^{-\epsilon} = 0$.

In \vampl\ and \scalar, the amplitude is expressed as a sum of
many dilogarithms -- five for every box function. However, the
first four dilogarithms in \scalar\ can be eliminated, in favor of
products of logarithms, when one performs the sum in \vampl. Thus,
the total amplitude can be alternatively written as a sum
involving only the fifth dilogarithm in \scalar\ (plus products of
logarithms). This simplified form for the amplitude was obtained
in \BernZX, and makes it possible to write the amplitude for
$n\leq 5$ in terms of logarithms only. However, it turns out that
to understand the twistor space structure of the amplitudes, it is
better to work directly with the box functions.

{}From the definition of the box function \scalar, it follows that
it is really just a function of three vectors.  Two of them are
$p_{i-1}$ and $p_{i+r}$, the momenta of the two external gluons
that are not combined.   Of course, these vectors are lightlike.
To simplify the following formulas, we call these vectors $p$ and
$q$. The third vector that enters the box function is the total
momentum $P=p_i+p_{i+1}+\dots+p_{i+r-1}$ of one set of combined
gluons. We also set $Q=-p-q-P=p_{i+r+1}+p_{i+r+2}+\dots +p_{i-2}$,
the momentum of the other set of combined gluons, so momentum
conservation is expressed as $p+q+P+Q=0$.

Using the new notation, we define a generic scalar function as
follows,
\eqn\scalk{\eqalign{  F(p,q,P) & = -{1\over \epsilon^2}\left[
(-(P+p)^2)^{-\epsilon} + (-(P+q)^2)^{-\epsilon}
-(-P^2)^{-\epsilon} -(-Q^2)^{-\epsilon}\right] \cr & + {\rm Li}_2
\left( {1-{P^2\over (P+p)^2}} \right) + {\rm Li}_2 \left(
{1-{P^2\over (P+q)^2}} \right) \cr & + {\rm Li}_2 \left(
{1-{Q^2\over (Q+q)^2}} \right) + {\rm Li}_2 \left( {1-{Q^2\over
(Q+p)^2}} \right) \cr & - {\rm Li}_2 \left( {1-{P^2 Q^2\over
(P+p)^2(P+q)^2}} \right) + {1\over 2}\log^2\left( {(P+p)^2\over
(P+q)^2} \right).}  }
The connection with $F_{n:r;i}^{\rm 2m~e}$ is achieved by simply
taking $p=p_{i-1},$ $ q=p_{i+r}$ and $P=p_i+\dots +p_{i+r-1}$.

\subsec{Decomposition of the Amplitude}

Our goal is to decompose $F(p,q,P)$ as a sum of functions whose
twistor transforms are localized on simple algebraic sets. This
will be analogous to the representation of tree amplitudes as a
sum of MHV tree diagrams, obtained in \CachazoKJ.  The
decomposition will not be manifestly Lorentz covariant. As in
\CachazoKJ, to make the decomposition, we must define a positive
chirality spinor $\lambda_P$ for an arbitrary momentum vector $P$,
 not necessarily lightlike. We do this by introducing an arbitrary negative
chirality spinor $\eta^{\dot a}$ (which we take to be the same for
all $P$) and setting
\eqn\offs{ \lambda_{P\,a} = P_{a\dot a}\eta^{\dot a}.}
We also adopt this definition for a lightlike vector $p$, i.e.,
$\lambda_{p\,a} = p_{a\dot a}\eta^{\dot a}$. This is compatible
with but more precise than the usual definition; usually, for
lightlike $p$,  $\lambda_p$ is defined up to scaling by requiring
that $p_{a\dot a}=\lambda_{p\,a}\tilde\lambda_{p\,\dot a}$ for
some $\tilde\lambda_{p\,\dot a}$.  The virtue of breaking this
scaling symmetry and using \offs\ for all momenta, lightlike or
not, is that it ensures identities like $\langle
p,P+q\rangle=\langle p,P\rangle +\langle p,q\rangle$.

We define the inner product
\eqn\inner{ \vev{P,Q} = \epsilon_{a b} \lambda_P^a\lambda_Q^b}
for any momenta $P$ and $Q$, lightlike or not.

With these definitions, we can factorize the argument of the fifth
dilogarithm in \scalk, using the following rather unexpected
identity:
\eqn\factor{ {1-{P^2 Q^2\over (P+p)^2(P+q)^2}} = {(1-x)(1-y)\over
x y}}
where
\eqn\defxy{ x = {\vev{p,P}\over \vev{p,P+q}}{(P+q)^2\over P^2}
,\qquad y = {\vev{q,P}\over \vev{q,P+p}}{(P+p)^2\over P^2}.}
This identity holds for any choice of $\eta$. To prove the
identity, note first of all that $\tilde\lambda_p^{\dot a}$ and
$\tilde\lambda_q^{\dot a}$ give a basis for the space of negative
chirality spinors, so we can assume that $\eta^{\dot
a}=\alpha\tilde\lambda_p^{\dot a}+\beta\tilde\lambda_q^{\dot a}$
for some scalars $\alpha$, $\beta$.  Moreover, if
$(\alpha,\beta)=(1,0)$, then \factor\ is straightforwardly
verified using $\langle p,P\rangle=2p\cdot P$, $\langle
p,P+q\rangle=2p\cdot(P+q)$, and $\langle q,P\rangle=\langle
q,P+p\rangle$.  So it suffices to show that the right hand side of
\factor\ is independent of $\alpha$ and $\beta$.  For  any vector
$P_{a\dot a}$ and spinors $\tau^a$, $\tilde\nu^{\dot a}$, we set
$\langle\tau|P|\tilde\nu]=\tau^aP_{a\dot a}\tilde\nu^{\dot a}$.
Then we compute \eqn\nnons{{1-x\over x}={1\over (P+q)^2}{\alpha(2p
\cdot q P^2-4P\cdot p P\cdot q)+\beta(-2P\cdot q\,\langle
p|P|q])\over 2\alpha p\cdot P+\beta\langle p|P|q]},} and likewise
\eqn\onons{{1-y\over y}={1\over (P+p)^2}{\beta(2p \cdot q
P^2-4P\cdot p P\cdot q)+\alpha(-2P\cdot p\,\langle q|P|p])\over
\alpha\langle q|P|p]+2\beta P\cdot q}.} Let us write
$(1-x)/x=(s\alpha+t\beta)/(u\alpha+v\beta)$ and
$(1-y)/y=(s'\alpha+t'\beta)/(u'\alpha+v'\beta)$, with coefficients
$s,t$, etc., that are independent of $\alpha$ and $\beta$.  We
claim that $(s\alpha+t\beta)/(u'\alpha+v'\beta)$ is independent of
$\alpha$ and $\beta$, as is $(s'\alpha+t'\beta)/(u\alpha+v\beta)$.
The claim clearly implies that $(1-x)(1-y)/xy$ is independent of
$\alpha$ and $\beta$, as desired.  For example, the condition for
$(s\alpha+t\beta)/(u'\alpha+v'\beta)$ to be independent of
$\alpha$ and $\beta$ is $sv'-tu'=0$.  In the present context, this
condition becomes \eqn\ucop{2p\cdot qP^2-4P\cdot p P\cdot
q+\langle p|P|q]\langle q|P|p]=0.} This identity -- which if the
spinors are written in conventional notation would be called a
Fierz identity -- holds for any lightlike vectors $p,q$ and any
vector $P$.  The other condition we need, namely $s'v-t'u=0$,
follows from the same identity.

The factorization \factor\ is useful because it enables us to use
Abel's dilogarithm identity:
\eqn\idtwo{\eqalign{ & {\rm Li}_2\left[ {(1-x)(1-y)\over x y}
\right] = \cr &   {\rm Li}_2\left( {1-x\over  y} \right) + {\rm
Li}_2 \left( {1-y\over x } \right) - {\rm Li}_2\left( 1-x \right)
- {\rm Li}_2\left( 1-y \right) -\log x \log y. }  }

We can get more insight by means of further identities analogous
to \factor.
 By momentum conservation
\eqn\symm{{1-{P^2 Q^2\over (P+p)^2(P+q)^2}} = {1-{P^2 Q^2\over
(Q+q)^2(Q+p)^2}}.}
So upon introducing the variables related to $x$ and $y$ by
exchange of $P$ and $Q$
\eqn\mirror{ {\tilde x} = {\vev{p,Q}\over
\vev{p,Q+q}}{(Q+q)^2\over Q^2} ,\qquad {\tilde y} =
{\vev{q,Q}\over \vev{q,Q+p}}{(Q+p)^2\over Q^2},}
we have
\eqn\conse{ {(1-x)(1-y)\over x y}={(1-\tilde x)(1-\tilde y)\over
\tilde x \tilde y}. }
In addition,
\eqn\comi{ {1-x\over y} = -{1-\tilde x \over \tilde x}, \qquad
{1-y\over x} = -{1-\tilde y \over \tilde y}.}
Only one of these two identities requires a proof since the other
one follows from the first and \conse.


The proof of \comi\ goes along the same lines as the proof of
\factor. First, the expression
\eqn\equi{ \left( {\tilde x\over 1-\tilde x}\right) \left(
{1-x\over y} \right)}
is readily shown to equal $-1$ if $(\alpha,\beta) =(1,0)$.  We
show that it is independent of $\alpha$ and $\beta$ as follows. We
use the definitions
\defxy\ and  \mirror, together with   momentum conservation, to find
\eqn\alre{ \left( {\tilde x\over 1-\tilde x}\right) \left(
{1-x\over y} \right) = \left( {\vev{p,P+q}P^2 -\vev{p,P}(P+q)^2
\over \vev{q,P}} \right) \left({\vev{q,Q}\over
\vev{p,Q+q}Q^2-\vev{p,Q}(Q+q)^2} \right) .}
The first factor on the right equals
$(s\alpha+t\beta)/(u'\alpha+v'\beta)$ times a factor that is
trivially independent of $\alpha$ and $\beta$ (here $s,t,u'$, and
$v'$ are as before), and so is independent of $\alpha$ and
$\beta$. The second factor is also independent of $\alpha$ and
$\beta$, since it can be obtained from the first by exchanging $P$
and $Q$.

Having proven that
\eqn\rewr{{1-x\over y} = 1-{1 \over \tilde x}, \qquad {1-y\over x}
= 1-{1 \over \tilde y},}
we can use Landen's identity\foot{This form of Landen's identity
is valid for $z>0$. For $z<0$ one has to add $- 2\pi i \log
(1-z)$.}
\eqn\idone{{\rm Li}_2(1-z) + {\rm Li}_2\left( 1-{1\over z}\right)
= -{1\over 2}\log^2(z)}
on ${\rm Li}_2\left( 1-x \right)$ and ${\rm Li}_2\left( 1-y
\right)$ in \idtwo\ to get
\eqn\defin{ \eqalign{ & {\rm Li}_2\left[ {(1-x)(1-y)\over x y}
\right] = \cr &   {\rm Li}_2\left( 1-{1\over \tilde x} \right) +
{\rm Li}_2 \left( 1-{1\over \tilde y } \right) + {\rm Li}_2\left(
1-{1\over x} \right) + {\rm Li}_2\left( 1-{1\over y} \right)
+{1\over 2}\log^2\left({x\over y} \right).} }
Note that by momentum conservation $x/y = \tilde y/\tilde x$ and
therefore the right hand side of
\defin\ is  symmetric under  exchanging  $P$ and $Q$
or  $p$ and $q$.

If we use \defin\ to re-express the fifth dilogarithm in  \scalk,
each of the four dilogarithms on the right hand side of \defin\
can naturally be grouped with one of the first four dilogarithms
already present in \scalk. We get an expression with eight
dilogarithms, of which two are
\eqn\boxa{-{\rm Li}_2 \left( 1 - {\vev{p,P+q}\over
\vev{p,P}}{P^2\over (P+q)^2} \right) + {\rm Li}_2 \left( 1 -
{P^2\over (P+q)^2} \right),}
while the others are obtained from \boxa\ by exchange of $p$ and
$q$ and and/or exchange of $P$ and $Q$.

It turns out that a particular combination of \boxa\ with some
terms involving products of logarithms gives a function whose
twistor transform is supported on a simple algebraic set. Such a
combination is
\eqn\local{\eqalign{ H_q (p,P) = &  - {\rm Li}_2 \left( 1 -
{\vev{p,P+q}\over \vev{p,P}}{P^2\over (P+q)^2} \right) + {\rm
Li}_2 \left( 1 - {P^2\over (P+q)^2} \right) \cr & + \log \left(
{\vev{p,P+q}\over \vev{p,P}} \right) \log \left( {(P+q)^2\over
\mu^2}\right) - {1\over 4}\log^2\left( {\vev{p,P+q}\over
\vev{p,P}} \right) .}}
where $\mu$ is an arbitrary scale. Introducing such a scale is
natural for the divergent terms in the box function but somewhat
unnatural for a function that contributes only to the finite part.
However, it is easy to check that in the combination $H_q(p,P) +
H_p(q,P) + H_q(p,Q) + H_p(q,Q)$ the $\mu$ dependence cancels.
Moreover, after some algebra, we find that the box function can be
written
\eqn\mainr{\eqalign{  F(p,q,P)  =& -{1\over \epsilon^2}\left[
(-(P+p)^2)^{-\epsilon} + (-(P+q)^2)^{-\epsilon}
-(-P^2)^{-\epsilon} -(-Q^2)^{-\epsilon}\right] \cr + ~& H_q(p,P) +
H_p(q,P) + H_q(p,Q) + H_p(q,Q)  - \log\left( {\vev{p,Q}\over
\vev{p,P}}\right) \log \left( {\vev{q,P} \over \vev{q,Q}}
\right).}}
As we will now explain, this formula gives a convenient way to
understand the twistor space support of the scattering amplitude.

\subsec{Twistor interpretation}

There are three different building blocks in the decomposition of
the box function \mainr. They are of the form
\eqn\fuke{ -{1\over \epsilon^2}(-P^2)^{-\epsilon}, \quad H_q(p,P),
\quad {\rm and} \quad \log\left( {\vev{p,Q}\over \vev{p,P}}\right)
\log \left( {\vev{q,P} \over \vev{q,Q}} \right).}

The twistor transform of $-{1\over \epsilon^2}(-P^2)^{-\epsilon}$,
where $P = p_i +p_{i+1} + \ldots + p_{j-1}+p_j$ for some $i$ and
$j$, is localized on two  lines $L$ and $L'$ that generically are
disjoint. That the gluons in the set $\{i,i+1\dots,j\}$ are
contained in a line $L$ is clear from the criterion stated at the
end of section 2.1: the amplitude only depends on the sum of their
momenta.  Likewise, the other gluons $\{j+1,j+2,\dots,i-1\}$ are
contained in a second line $L'$.   Generically, $L$ and $L'$ do
not intersect; they are contained in no common plane (see fig.
5(a)). This can be checked by using the operator $K_{klmn}$
defined in section 2, with $k,l$ corresponding to two gluons in
$L$ and $m,n$ two gluons in $L'$. It is not difficult to prove by
hand that no power of this operator annihilates $-{1\over
\epsilon^2}(-P^2)^{-\epsilon}$.

\bigskip
\ifig\hula{Twistor configurations contributing to the box
function. (a) Two disjoint lines. (b) Two intersecting lines with
$q$ in the plane. Hypothetical twistor space propagators are not
shown.}
 {\epsfxsize=0.75\hsize\epsfbox{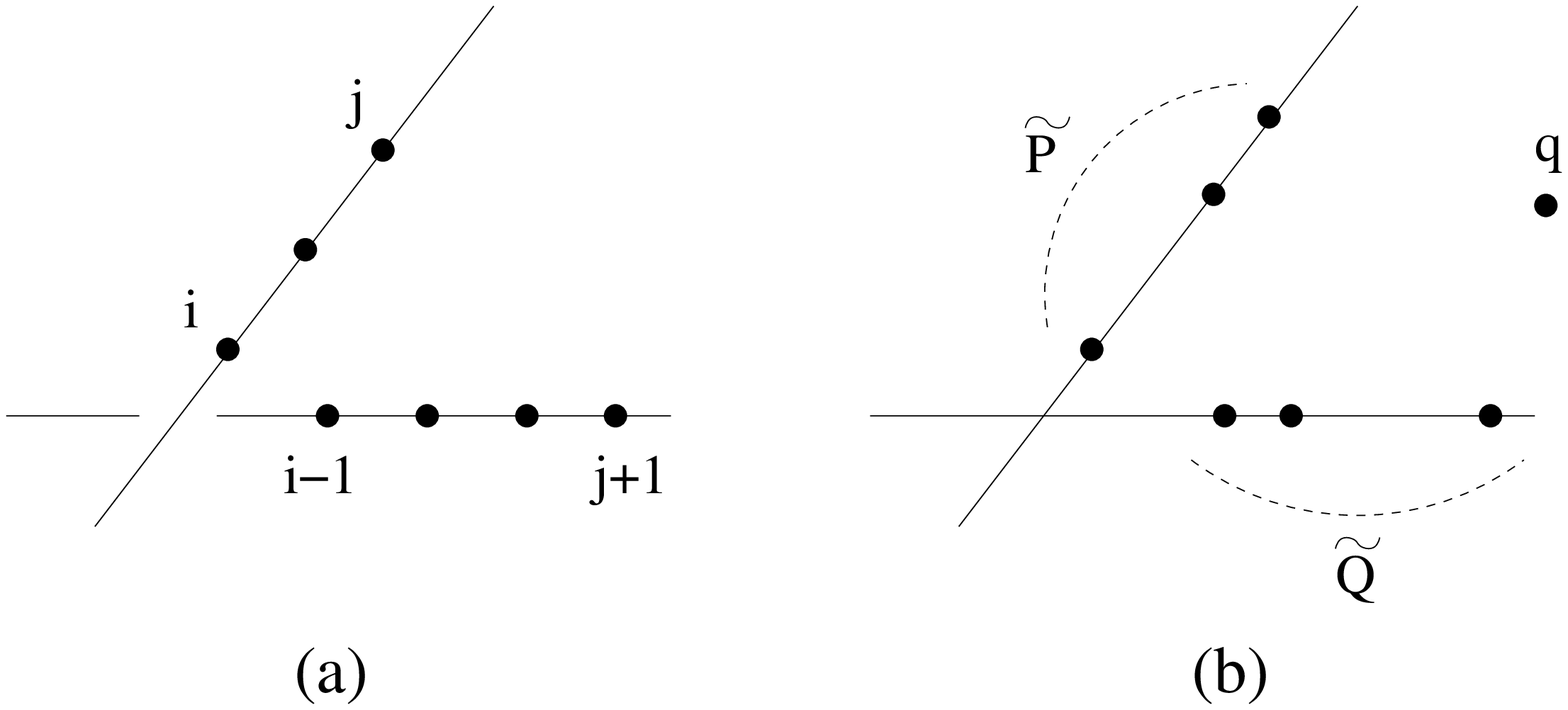}}
\bigskip

We now turn to $H_q(p,P)$. We write $\tilde P$ for the set of
gluons whose momenta adds to $P$, and $\tilde Q$ for  the set of
gluons whose momenta adds to $Q+p$.   The twistor transform of
$H_q(p,P)$ is localized on a configuration in which gluons in
$\tilde P$ are contained in one line $L$, while gluons in $\tilde
Q$ are contained in another line $L'$. Moreover, the two lines $L$
and $L'$ intersect and so lie on a plane. The remaining gluon $q$
is contained in this plane (with  derivative of  delta function
support). See \hula(b).

The statement about the collinearity of gluons in $\tilde P$, and
likewise for collinearity of gluons
 in $\tilde Q$, follows from the criterion stated
at the end of section 2.1. Indeed, the dependence of the amplitude
on gluons in $\tilde P$ is only via the sum of the momenta; the
dependence on gluons in $\tilde Q$ is only via the holomorphic
spinor $\lambda_p$  and the sum of the momenta.

That the two lines $L$ and $L'$ intersect is a much less trivial
fact. Indeed, appendix B is devoted to a proof of it. The proof is
made by showing that any set of two points  $P_1,P_2\in \tilde P$
and two points $Q_1,Q_2\in \tilde Q$ are coplanar. This is tested
by using the coplanar operator and showing that
\eqn\teti{ K_{P_1,P_2,Q_1,Q_2} H_q(p,P) = 0.}

It is also true that $q$ is contained (with derivative of delta
function support) in the plane containing $L$ and $L'$.  To show
this, one has to check that $H_q(p,P)$ is annihilated by $K^2$,
that is by any product $K_{ijkl}K_{i'j'k'l'}$. If any of the two
$K$'s does not contain $q$, then already a single $K$ annihilates
the amplitude.  ($K_{ijkl}H_p(p,P)=0$ according to \teti\ if two
of the points $i,j,k,l$ are in $\tilde P$ and two in $\tilde Q$;
if three or more are contained in $\tilde P$ or $\tilde Q$, it
vanishes because three collinear points and any fourth point lie
in a plane.) So the key case is that $q$ is contained in
$\{i,j,k,l\}$ and also in $\{i',j',k',l'\}$. We have not been able
to find an analytic proof that $K_{ijkq}K_{i'j'k'q}H_q(p,P)=0$.
However, we have verified this with computer assistance for up to
seven gluons.

Finally, consider the remaining logarithmic term in \mainr:
\eqn\huli{ \log\left( {\vev{p,Q}\over \vev{p,P}}\right) \log
\left( {\vev{q,P} \over \vev{q,Q}} \right).}
The twistor transform of \huli\ is localized on a plane. Indeed,
any function of only the inner products $\vev{\lambda , \lambda'}$
of positive chirality spinors (of on-shell and off-shell momenta)
is localized on a plane. One way to see this is again by showing
that any operator $K_{ijkl}$ annihilates it. From the definition
of $K$ in section 2, it is schematically given by $K =
\lambda^{1}\lambda^{2}\mu^{\dot 1}\mu^{\dot 2}$. Therefore, under
the twistor transform $\mu \rightarrow i\del/\del \tilde\lambda$,
$K$ becomes a differential operator of degree two such that each
term contains one derivative with respect to the
$\tilde\lambda^{\dot 1}$ component of a gluon and one with respect
to  the $\tilde\lambda^{\dot 2}$ component of another gluon. Then
for $\eta^{\dot a} = \delta^{\dot a \dot 2}$, it is easy to see
that any function of inner products $\vev{ \lambda,\lambda'}$ is
independent of the $\tilde\lambda_{\dot 1}$ component of all the
gluons, and so is annihilated by $K$.  Obviously, this conclusion
does not depend on the choice of $\eta$.

\ifig\lopu{Twistor support of (a) $\log\vev{p,Q}\log\vev{q,Q}$,
(b) $\log\vev{p,P}\log\vev{q,P}$, (c)
$\log\vev{p,Q}\log\vev{q,P}$, and (d)
$\log\vev{p,P}\log\vev{q,Q}$.}
{\epsfxsize=0.75\hsize\epsfbox{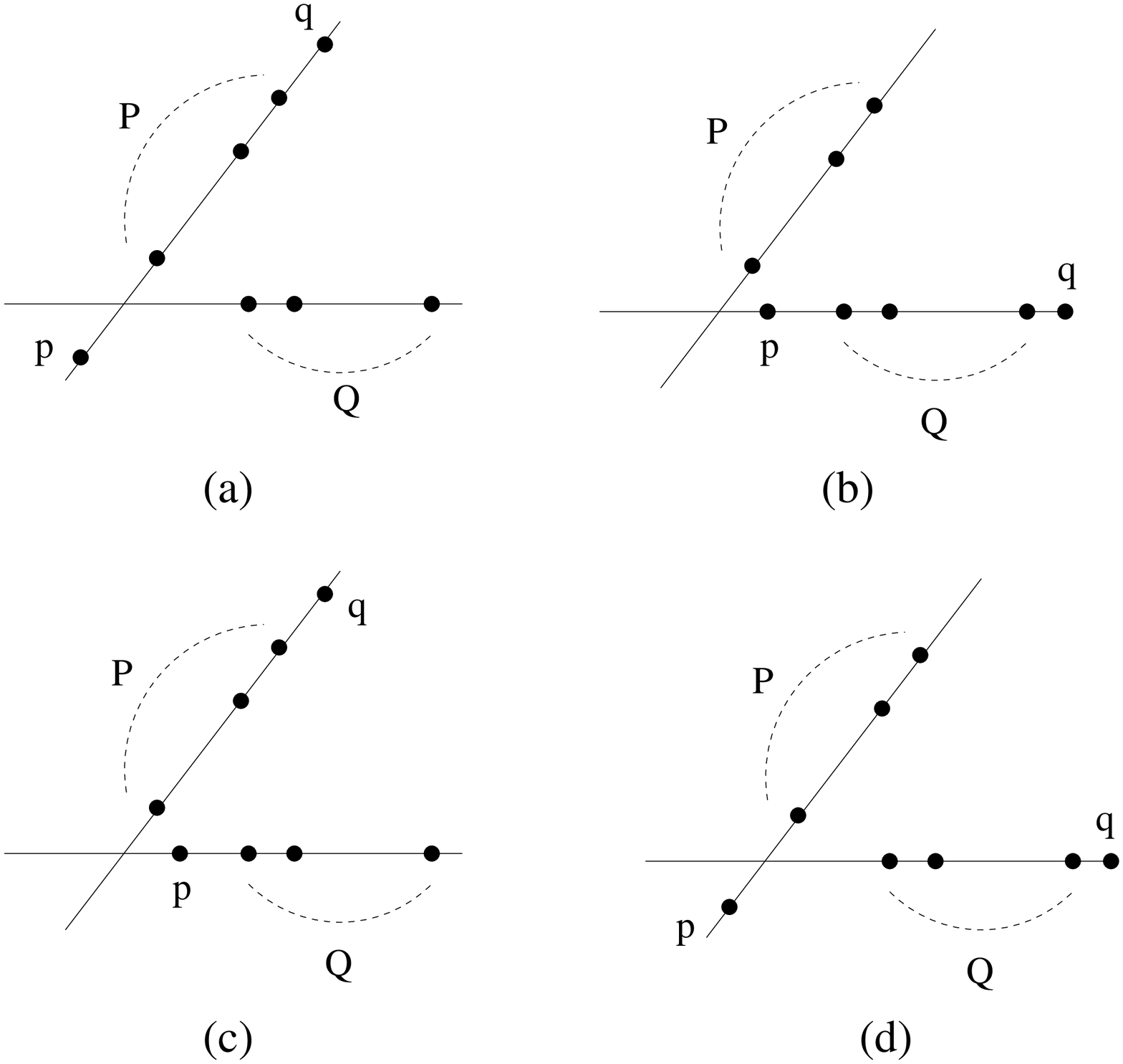}}

But more is true. Expanding \huli, we find four terms such as
$\log\vev{p,Q}\log\vev{q,Q}$. Each of them is localized on two
intersecting lines. The precise distribution of gluons can be
deduced from the criterion at the end of section 2.1 and is shown
in \lopu.

\subsec{Concluding Remarks}

So we have determined the twistor space structure of the one-loop MHV amplitudes
with ${\cal N}=4$ supersymmetry.  But a few remarks are in order.

The decomposition that we have made is good enough to determine
the twistor space structure of the amplitude, but a natural
evaluation of these amplitudes in a suitable twistor-string theory
might lead to a somewhat different decomposition.  For example,
the definition of $H_q(p,P)$ could be modified by adding some
logarithmic terms.

The twistor transform of $H_q(p,P)$ describes two collinear sets
of gluons, $\tilde P$ and $\tilde Q$, with an exceptional gluon
(of momentum $p$) in the set $\tilde Q$.  Obviously, in another
box function contributing to the scattering amplitude \vampl,
decomposed in the same fashion, there is a contribution with the
same collinear sets $\tilde P$ and $\tilde Q$, but the exceptional
gluon contained in $\tilde P$.  In a suitable twistor-string
computation, a single diagram might give the sum of these two
contributions.

After removing the infrared-divergent part, which is supported on
a pair of disjoint lines in twistor space, we have written the infrared-finite
one-loop amplitude as a sum over subamplitudes (essentially the
$H_q(p,P)$) associated with choices of how to combine a subset of
adjacent gluons. Let ${\cal S}$ be the set of such choices.  For
each $s\in {\cal S}$, let ${\cal D}_s$ be one of the differential
operators (a suitable $F_{ijk}$ or a suitable power of some
$K_{ijkl}$) that annihilates this subamplitude. Then
our decomposition of the one-loop infrared-finite MHV amplitudes makes manifest that these
amplitudes are annihilated by the differential operators $\prod_{s\in {\cal S}}{\cal D}_s$.
A considerable amount of computer-based inquiry indicates that no other
products of $F$'s and $K$'s annihilate these amplitudes.  Thus, adding up the subamplitudes does
not lead to any further simplification of the twistor space structure.

See Appendix C for a discussion of a covariant decomposition of the box function.

\newsec{${\cal N}=1$ One-Loop MHV Amplitude}

The remainder of this paper is devoted to an analysis of (some)
one-loop amplitudes with external gluons and internal massless
particles in theories with reduced supersymmetry.  The internal
particles in the loop may have spin $0$, $1/2$, or 1. A convenient
basis for these three one-loop amplitudes is to consider the
${\cal N}=4$ amplitude $A^{{\cal N}=4}$, which was the subject of
section 3, the amplitude $A^{{\cal N}=1}_{ch}$ due to an ${\cal
N}=1$ chiral multiplet, which will be the subject of this section,
and the amplitude $A_{sc}$ with a scalar in the loop, which we
consider in section 5.

 \subsec{${\cal N}=1$ MHV Amplitude}

${\cal N}=1$ supersymmetry -- like ${\cal N}=4$ -- leads to the
vanishing of gluon scattering amplitudes in which all external
gluons or all but one have the same helicity.  The first and
simplest amplitude is thus the MHV amplitude with precisely two
gluons of negative (or positive) helicity.  These amplitudes have
been computed for any number $n$ of external gluons.  The
computations are made \BernCG\ by expressing the gluon scattering
amplitudes in terms of scalar one-loop integrals with two, three,
or four external lines; these are sometimes called bubble,
triangle, or box integrals.  We review the formulas here and then
describe the twistor space structure in section 4.2.

In contrast to the ${\cal N}=4$ one-loop MHV amplitudes, the
analogous ${\cal N}=1$ amplitudes depend nontrivially on which
gluons have negative helicity.  For  an $n$-gluon process in which
gluons $i$ and $j$ have negative helicity, the amplitude is
\BernCG\
 \eqn\nofu{\eqalign{A^{{\cal
N}=1}_{ch}&=A^{\rm tree}\times \biggl\{ \sum_{\p=i+1}^{j-1}
\sum_{\q=j+1}^{i-1} b_{\p,\q}^{i,j}
B(t_{\p+1}^{[\q-\p]},t_\p^{[\q-\p]};t_{\p+1}^{[\q-\p-1]},t_{\q+1}^{[\p-\q-1]})
\cr &+ \sum_{\p=i+1}^{j-1} \sum_{a=j}^{i-1 } c^{i,j}_{\p,a}
{\ln(t_{\p+1}^{[a-\p]}/t_\p^{[a-\p+1]}) \over t_{\p+1}^{[a-\p]}-
t_\p^{[a-\p+1]}} + \sum_{\p=j+1}^{i-1} \sum_{a=i}^{j-1}
c_{\p,a}^{i,j}{\ln(t_{a+1}^{[\p-a]}/t_{a+1}^{[\p-a-1]}) \over
t_{a+1}^{[\p-a]}-t_{a+1}^{[\p-a-1]}} \cr &+{c_{i+1,i-1}^{i,j}\over
t_{i}^{[2]}} K_0(t_i^{[2]})+{c_{i-1,i}^{i,j}\over t_{i-1}^{[2]}}
K_0(t_{i-1}^{[2]})+{c_{j+1,j-1}^{i,j} \over t_j^{[2]}}
K_0(t_j^{[2]})+{c_{j-1,j}^{i,j} \over t_{j-1}^{[2]}}
K_0(t_{j-1}^{[2]}) \biggr\}.}} Here $t_i^{[k]}=(p_i+p_{i+1}+\dots
+p_{i+k-1})^2$ for $k\geq 0$, and $t_i^{[k]}=t_i^{[n-k]}$ for
$k<0$.  Sums are taken in cyclic order around the circle, so a sum
$\sum_{k=i}^j$ is evaluated by summing over all $k$ in the
clockwise direction from $i$ to $j$, regardless of whether $i$ is
greater than or less than $j$. Though we have not indicated this
in writing the formula, the sum over $a$ and $\p$ is restricted to
$|a-\p|>1$, $|a+1-\p|>1$, so that the logarithms have a finite,
nonzero argument.

The coefficients in front of the integral functions are
\eqn\coeffn{\eqalign{b_{\p,a}^{i,j}&=2{\vev{i,\p}\vev{\p,j}\vev{i,a}\vev{a,j}
\over \vev{i,j}^2 \vev{\p,a}^2}\cr
c_{\p,a}^{i,j}&={(\tr_+[\slashed{p}_i \slashed{p}_j \slashed{p}_\p
\slashed{q}_{\p,a}]-\tr_+[\slashed{p}_i\slashed{p}_j\slashed{q}_{\p,a}\slashed{p}_\p])
\over (p_i+p_j)^2} {\vev{i,\p}\vev{\p,j}\over \vev{i,j}}
{\vev{a,a+1} \over \vev{a,\p} \vev{\p,a+1}},}} where we have set
$q_{i,j}=\sum_{l=i}^{j}p_l$ and where
\eqn\ttp{\eqalign{\tr_+[\slashed{p}_{a_1}\slashed{p}_{a_2}
\slashed{p}_{a_3} \slashed{p}_{a_4}]&=\half \tr [(1+\gamma_5)
\slashed{p}_{a_1} \slashed{p}_{a_2} \slashed{p}_{a3}
\slashed{p}_{a4}]=[a_1a_2] \vev{a_2a_3}[a_3a_4]\vev{a_4a_1} .}}

\ifig\twoms{The scalar box integral contributing to the amplitude.
Two of the vertices carry light-like momenta $p$ and $q$. $P$ and
$Q$ are sums of several light-like momenta. One negative-helicity
gluon is in $P$ and one is in $Q$; we label them as $i\in P$ and
$j\in Q.$ } {\epsfxsize=0.35\hsize\epsfbox{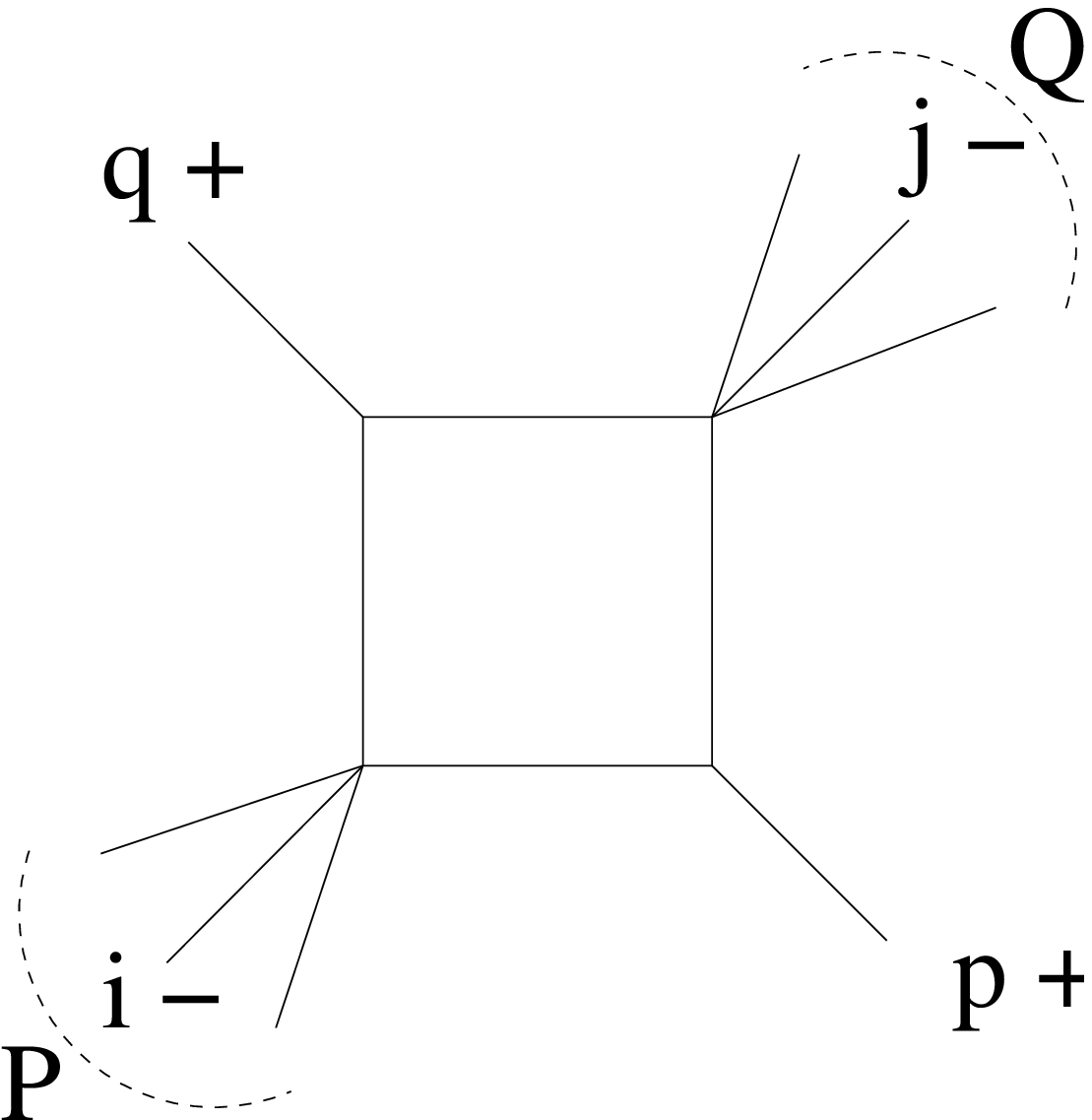}}

The function $B$ comes from the scalar one-loop integral with two
masses, shown in  \twoms\ and also discussed in Appendix A. Using
the conventions shown in the figure, $p=p_\p$ and $q=p_\q$ while
$P=p_{\p+1}+p_{\p+2}+\dots+p_{\q-1}$ and
$Q=p_{\q+1}+p_{\q+2}+\dots+p_{\p-1}$.  So $p,q,P,Q$ are the four
incoming momenta of the box diagram. The scalar function $B$ is
the finite part\foot{This is a slight imprecision in our  language
here since the divergent terms, proportional to $1/\epsilon^2$, in
\scalk\ also contain finite pieces which we do not include in
$B$.} of the ${\cal N}=4$ scalar box function \scalk\ studied in
section 3: \eqn\moto{\eqalign{B(p,q,P,Q)&=F^{\rm
finite}(p,q,P,Q)\cr &={\rm Li}_2\left(1-{P^2 \over (P+p)^2}\right)
+{\rm Li}_2\left(1-{P^2 \over (P+q)^2}\right)\cr &+{\rm
Li}_2\left(1-{Q^2 \over (Q+q)^2}\right)+{\rm
Li}_2\left(1-{Q^2\over(Q+p)^2}\right)\cr &-{\rm
Li}_2\left(1-{P^2Q^2 \over (P+p)^2
(P+q)^2}\right)+\half\log^2\left((P+p)^2\over(P+q)^2\right).}}

\ifig\traok{Triangle diagram contributing to the amplitude. $p$ is
a lightlike momentum, $P$ is a sum of light-like momenta
containing $i$ and $Q$ is a sum of momenta containing $j$.}
{\epsfbox{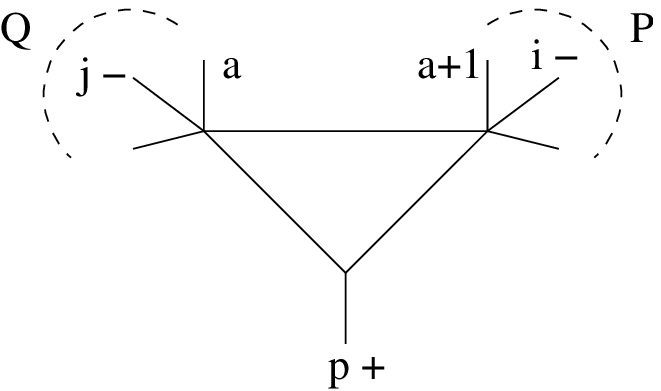}}

The logarithmic terms in  \nofu\ arise from a scalar triangle
diagram that is indicated in \traok.\   It is convenient to write
$p=p_m$ and $Q$ and $P$ for the sums $p_{\p+1}+p_{\p+2}+\dots+p_a$
and $p_{a+1}+p_{a+2}+\dots + p_{\p-1}.$ One of these sums contains
$p_i$ and one contains $p_j$; we write $P$ for the sum that
contains $p_i$ and $Q$ for the sum that contains $p_j.$ Using the
variables $p,P,$ and $Q$, we can rewrite the complicated-looking
expressions such as
\eqn\jjno{{\ln(t_{\p+1}^{[a-\p]}/t_\p^{[a-\p+1]}) \over
t_{\p+1}^{[a-\p]}- t_\p^{[a-\p+1]}}} in the convenient form
\eqn\trif{T(p,P,Q)={\ln(Q^2/P^2) \over Q^2-P^2}} The coefficient
$c_{\p,a}^{i,j}$ in \coeffn\ can be simplified using the
definition \ttp:
\eqn\capas{\eqalign{c_{\p,a}^{i,j}&={\vev{i,\p}\vev{\p,j} \over
\vev{i,j}^2}{\vev{a,a+1} \over \vev{a,\p}\vev{a,\p+1}} \cr
&\times\cases{ \bigl( \vev{j,\p} \langle i|P|\p]+\vev{i,\p}\langle
j|P|\p] \bigr),& $\p=j+1,\dots,i-1$ \cr  \bigl(\vev{j,\p}\langle
i|Q|\p]+ \vev{i,\p}\langle j|Q|\p] \bigr), & $\p=i+1,
\dots,j-1$.\cr}\cr}} The main feature that we will use in
remainder of the discussion is that the antiholomorphic dependence
of the coefficients \capas\ (that is, their dependence on the
negative chirality spinors $\tilde\lambda$) is captured in $p,P$
and $Q.$ In particular, these coefficients are holomorphic in
$i,j,a.$

The amplitude \nofu\ diverges when gluon $i$ or $j$ becomes
collinear with one of the adjacent positive helicity gluons, which
we will label $g$. For $g=i-1$ or $g=i+1$, the piece that diverges
when $p_i$ and $p_g$ become collinear can be evaluated in terms of
a scalar bubble diagram that depends on $P=p_i+p_g$. It can be
simplified to \eqn\disi{{c_{g,a}^{i,j}\over s_{i,g}} K_0(s_{i,g})=
-{\vev{i,g}\vev{g,j} \over \vev{i,j}} {\vev{a,a+1}\over \vev{a,g}
\vev{g,a+1}}{1\over \epsilon(1-2\epsilon)}(-P^2)^{-\epsilon},}
where $a=i,i-1$ for $g=i-1,i+1$ respectively.  These terms are
infrared-divergent, as is evident from the pole at $\epsilon=0$;
we write them as $A_{IR}$.

\bigskip
\ifig\buba{The scalar bubble diagram giving the divergent part
$K_0(P^2)$ of the amplitude.}
{\epsfxsize=0.60\hsize\epsfbox{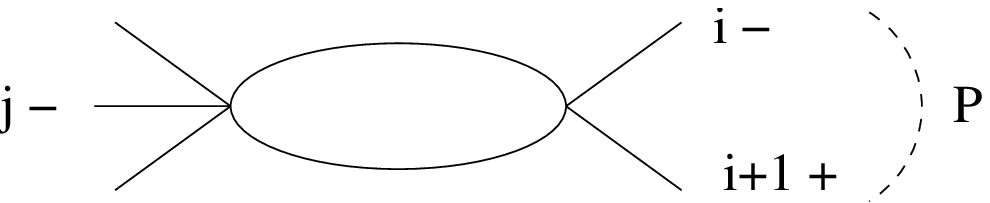}}

Collecting the different pieces, we can write the amplitude
schematically as the sum of box, triangle and bubble
contributions: \eqn\nosi{A^{N=1}_{ch}=A^{\rm tree}\times \left(
\sum_{\p,\q;i\in P,j\in Q} b^{i,j}_{\p,\q} B(p_\p,p_\q,P,Q)+
\sum_{\p,a;i\in P,j\in Q}c_{\p,a}^{i,j}T(p_\p,P,Q)+A_{IR}\right).}

\subsec{Interpretation}

\noindent{\it Box Diagrams}

We first discuss the contribution to the amplitude from the scalar
box functions, \eqn\obox{b_{\p,\q}^{i,j}B(p,q,P,Q).} Recall that
we have introduced $p=p_\p$ and $q=p_\q$ in order to simplify the
notation. The coefficient $b_{\p,\q}^{i,j}$ is a holomorphic
function, that is a function only of the $\lambda$'s, so it does
not affect the twistor space structure of the amplitude. Hence,
the localization of the box diagrams is determined by the the box
function $B(p,q,P,Q)$. This is the finite part of the scalar box
function \scalk, whose twistor-space structure was analyzed in
section 3. It corresponds to twistor-space configurations in which
the gluons whose momenta add to  $P$ are contained in one line $L$
while the gluons whose momenta add to  $Q$ are contained in
another line $L'$ that is coplanar with $L$; moreover, of the
remaining gluons $p$ and $q$, one is contained in $L$ or $L'$ and
one is contained in the plane containing $L$ and $L'$.

There are some interesting differences between the two cases.  For
the ${\cal N}=1$ chiral amplitude, of the two negative helicity
gluons $i$ and $j$, one is in $L$ and one is in $L'$.  (This
follows from the details of the sum in \nofu.) For ${\cal N}=4$,
there is no such restriction: any two gluons may have negative
helicity. Hence, for ${\cal N}=1$, there is always precisely one
negative helicity gluon on $L$ and one on $L'$, while for ${\cal
N}=4$, it is possible for both of these gluons to be on $L$ or
$L'$, or for one to be on $L$ or $L'$ and the other to be in the
bulk.  We will give an intuitive explanation of this in section
4.3, after considering the other contributions to the chiral
amplitude.

\bigskip\noindent
{\it Triangle Diagrams}

Similarly, we can see part of the twistor space structure of the
triangle amplitude  \eqn\tricc{c_{\p,a}^{i,j}T(p,P,Q)} without any
additional work, using the criterion stated at the end of section
2.1. As the $\tilde\lambda$'s only enter $c_{\p,a}$  via $p$, $P$,
and $Q$, the gluons whose momenta add to $P$ are supported on a
line $L$, and the gluons whose momenta add to $Q$ are supported on
another line $L'$.

Furthermore, all gluons are contained in a plane, that is a $\Bbb
{CP}^2$, just as in the ${\cal N}=4$ case.  In fact, we found with
some computer assistance that for up to seven gluons this
amplitude is annihilated by $K^2$, that is by any product
$K_{ijkl}K_{i'j'k'l'}$ of two collinear operators. The details of
the resulting ``derivative of a delta function support'' on
coplanar configurations are somewhat complicated, as one also has
\eqn\sop{K_{PPQQ}F_{pPP}F_{pQQ} \bigl( c_{\p,a}^{i,j}T(p,P,Q)
\bigr) =0.} Here, $K_{PPQQ}$ represents a coplanar operator
$K_{ijkl}$ with $i,j\in P$ and $k,l\in Q.$ Similarly $F_{pPP}$, or
$F_{pQQ}$, is a collinear operator $F_{pij}$ with $i,j\in P,$ or
in $Q$, respectively. This means roughly that while it is possible
to have a first order fluctuation away from coplanarity (since the
triangle amplitude is annihilated by $K^2$ but not by $K$), either
the two lines are strictly coplanar or one of them contains the
point $p$.

\bigskip\noindent
{\it Divergent Part}

Just as in the ${\cal N}=4$ case, the infrared divergent part of
the amplitude \nofu\ is \eqn\didi{{1\over \epsilon(1-2\epsilon)}
(-P^2)^{-\epsilon}} times a holomorphic function of spinors. As we
discussed in section 3, it localizes on a disjoint union of lines.
The gluons whose momenta add to $P$ are on one line and the
remaining gluons are on the second line.

\subsec{Comparison of Amplitudes}

The surprising result of sections 3 and 4.2 is that at one-loop
order, the ${\cal N}=4 $ MHV amplitude  and the ${\cal N}=1$
chiral MHV amplitude have very similar twistor space structure.
(We will see in section 5.3 that the cut-constructible part of the
nonsupersymmetric $--+++\dots +$ amplitude also has a similar
stucture.) In each case, apart from an elementary, infrared
divergent contribution that localizes on two disjoint lines, we
have found a novel twistor structure in which $n-1$ gluons are
contained in a pair of intersecting lines, and the remaining gluon
is in the plane that contains the lines.

\ifig\tini{A twistor configuration contributing to the $N=1$
chiral amplitude. All gluons except one are contained in a pair of
intersecting lines; the last gluon is in the plane containing the
lines.  We suppose that this last gluon is connected to the lines
by a twistor propagator, shown as a dashed line.  The gluons $i$
and $j$ are localized on opposite lines, because only scalar and
fermions can run around the loop. }
{\epsfxsize=0.45\hsize\epsfbox{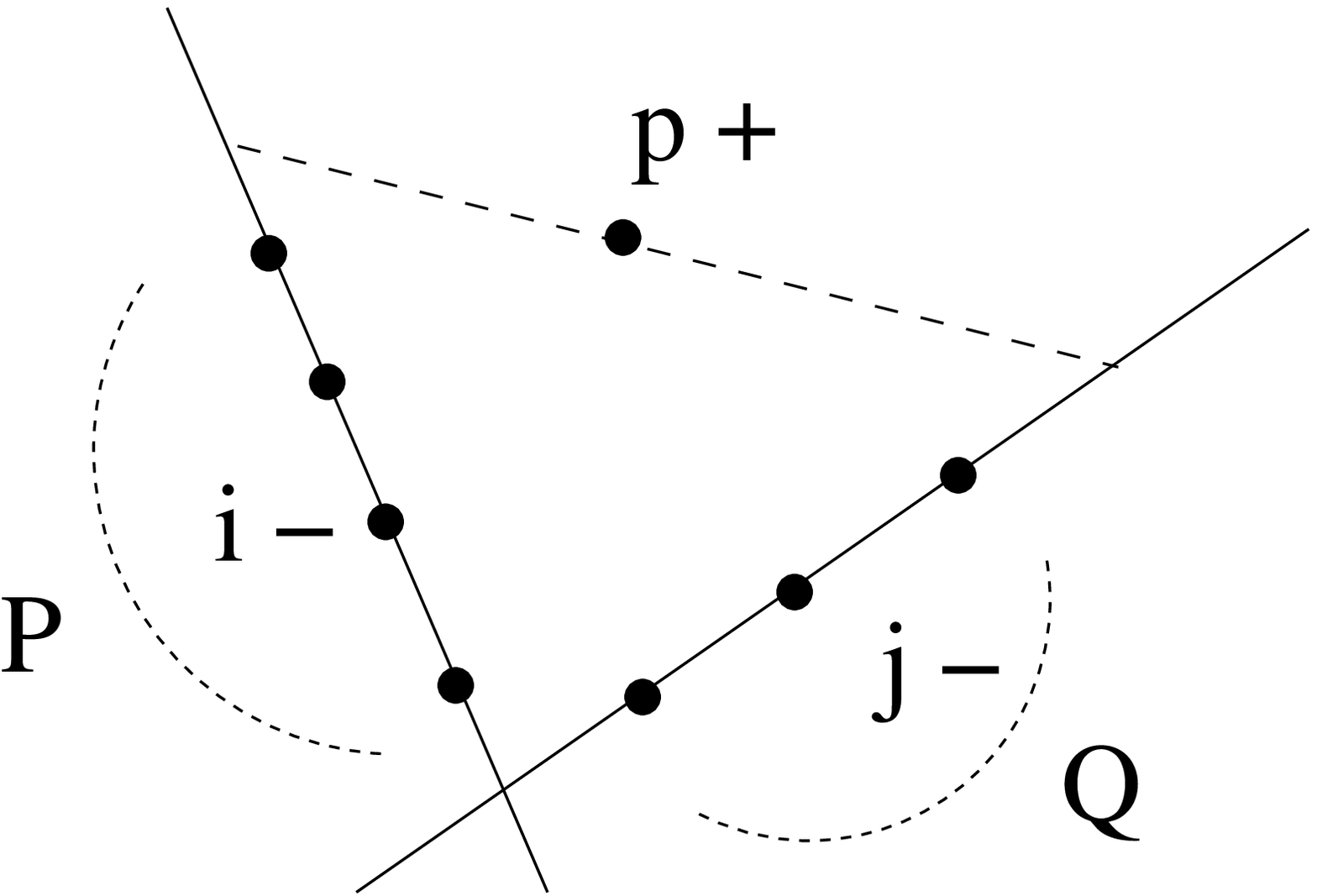}}

Unfortunately, we do not know how a twistor-string theory would
generate this structure.  A guess is indicated in \tini. (This
discussion needs to be revisited in view of a holomorphic anomaly
that will be discussed elsewhere.) Here we imagine two lines $L$
and $L'$ -- understood as $D$-instantons -- which intersect.
There is also a twistor space propagator connecting $L$ and $L'$.
One of the $n$ gluons is attached to this propagator.  We do not
know why $L$ and $L'$ intersect or why the gluon attached to the
propagator is contained in the same plane that contains $L$ and
$L'$.  The lines $L$ and $L'$ together with the twistor propagator
form a loop. Propagating around this loop is a particle of spin
$1,1/2,$ or $0$, in the case of the ${\cal N}=4$ amplitude, or
$1/2$ or 0, in the case of the ${\cal N}=1$ chiral amplitude.

In this picture, the $D$-instantons are of degree one and
represent MHV tree amplitudes, while the attachment of the
$n^{th}$ gluon to the twistor space propagator is presumably made
using a local twistor space interaction (similar to the ${\cal
A}^3$ twistor space interaction that comes from the Chern-Simons
form \WittenNN) . From this picture, we can see why for ${\cal
N}=4$, the negative helicity gluons can be inserted anywhere,
while for ${\cal N}=1$ one of them is inserted on $L$ and one on
$L'$.  In fact, for ${\cal N}=1$, the field propagating around the
loop has helicity $\leq 1/2$.  In a local, cubic interaction in
twistor space, the three helicities add up to $+1$, so such an
interaction does not couple a negative helicity gluon to fields of
helicity $\leq 1/2$. This explains why for ${\cal N}=1$, a
negative helicity gluon is not attached to the propagator in
\tini.  The two negative helicity gluons cannot be attached to the
same line $L$ or $L'$ for a similar reason: in a degree one tree
level amplitude with $k$ external fields of spin $1,1/2$, or 0,
the helicities must add up to $k-2$, which is not possible if two
particles have helicity $-1$ and one has helicity $\leq 1/2$.

\newsec{Nonsupersymmetric One-Loop Amplitudes}

In this section, we explore the twistor structure of some
nonsupersymmetric one-loop amplitudes.  The amplitudes that we
consider are scalar amplitudes, that is amplitudes with a massless
scalar propagating in the loop.  (As explained at the beginning of
section 4, nonsupersymmetric amplitudes due to a massless field of
spin $1/2$ or 1 are linear combinations of the scalar amplitudes
with supersymmetric amplitudes that we have already studied.)

A conspicuous difference between supersymmetric amplitudes and
nonsupersymmetric ones is that in the nonsupersymmetric case,
there exist $n$-gluon amplitudes in which $n$ or $n-1$ gluons have
positive helicity.  These amplitudes have been computed, and we
begin with them.  Then we consider the ``MHV'' amplitudes -- a
term which is not quite right in the nonsupersymmetric case --
with $n-2$ gluons of positive helicity.  These amplitudes are
known less completely.

\subsec{All Plus Helicity One-Loop Amplitudes}

The one-loop amplitude for $n\geq 4$ gluons all of positive
helicity is \refs{\bernplusa, \bernplusb} \eqn\allplus{A_n^{1-{\rm
loop}}(+,\dots,+)=-{i \over 48 \pi^2} \sum_{1\leq
i_1<i_2<i_3<i_4\leq n}{ \vev{i_1,i_2} [i_2,i_3] \vev{i_3,i_4}
[i_4,i_1] \over \vev{1,2} \vev{2,3}\cdots \vev{n,1}}.} We can also
write the amplitude in terms of the momenta and positive chirality
spinors of external particles \eqn\allplusb{A_n =-{i \over 96
\pi^2} \sum_{1\leq i_1<i_2<i_3<i_4\leq n} { s_{i_1 i_2} s_{i_3i_4}
-s_{i_1i_3}s_{i_2i_4}+s_{i_1i_4}s_{i_2i_4}-4i
\epsilon_{\mu\nu\lambda\rho}p_{i_1}^\mu p_{i_2}^\nu p_{i_3}^\rho
p_{i_4}^{\sigma}\over  \vev{1,2}\vev{2,3}\dots\vev{n,1}}.} This
amplitude is  single-valued and  free of cuts. Indeed, the
discontinuities of the all-plus one-loop amplitude across a cut
would be proportional to a product of tree amplitudes, at least
one of which would have less than two gluons of negative helicity
and so would vanish.

The twistor structure of this amplitude is clear.  The amplitude
is polynomial in the $\tilde\lambda$'s, so it is supported (with
derivative of a delta function support) on a line, that is, all
gluons are contained in some $\Bbb{CP}^1$. A line in twistor space
corresponds to a point in Minkowski spacetime.  So, like the tree
level MHV amplitude, the all-plus amplitude is a rough twistor
space analog of a local interaction.

\ifig\figtor{(a) A diagram with tree level MHV vertices.  Each
vertex couples to two negative helicity gluons; each propagator
absorbs one such gluon. (b) A diagram that also contains a
one-loop all-plus vertex, shown as a solid disc to symbolize the
fact that a loop is
hidden.}{\epsfxsize=0.55\hsize\epsfbox{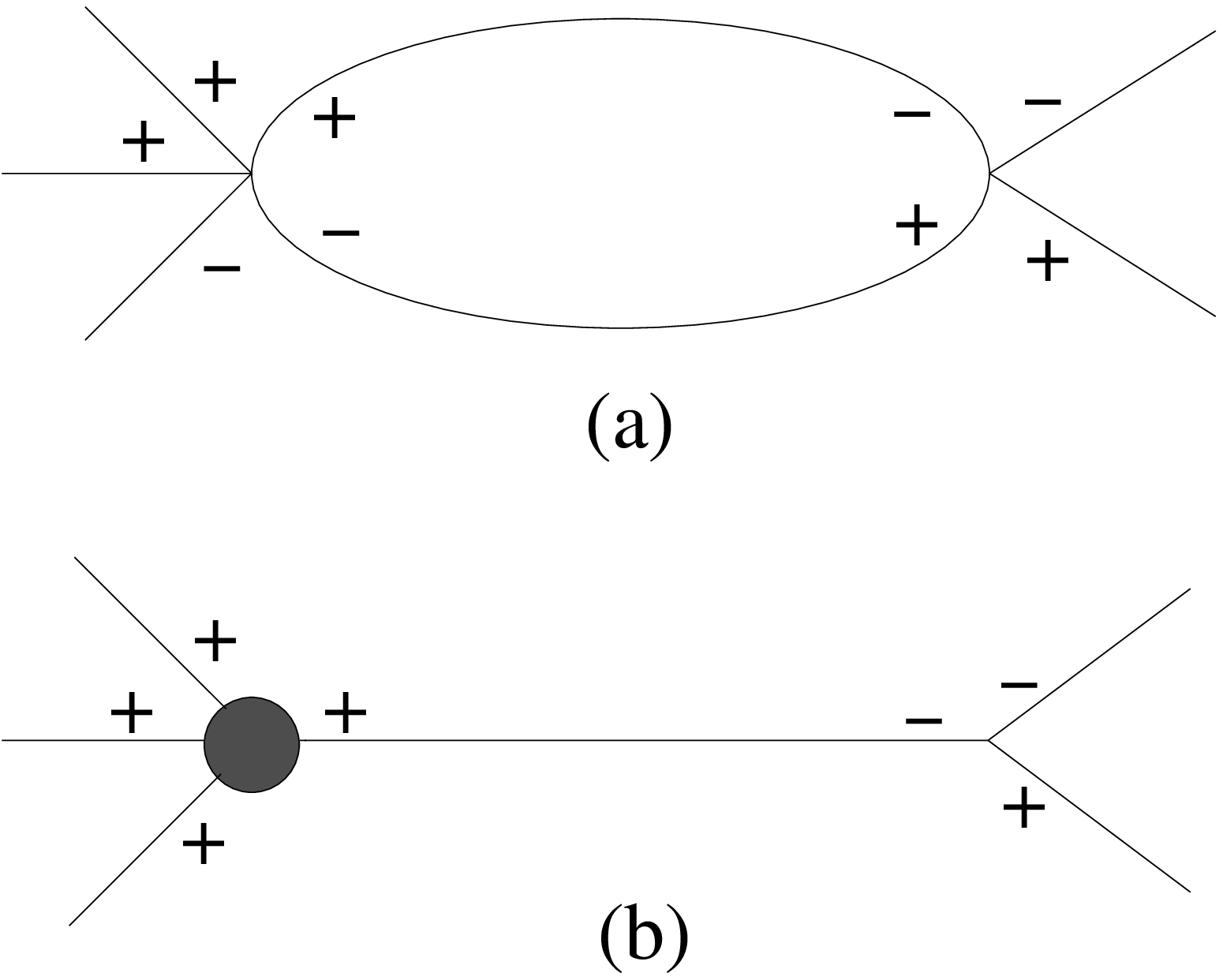}}

 It is consequently tempting to guess that one should extend the concept of
MHV tree diagrams \CachazoKJ\ to add the one-loop all-plus
amplitudes as new interaction vertices, in addition to the
familiar tree level MHV vertices. Before testing this idea out in
examples, let us make a few remarks about hypothetical Feynman
diagrams with vertices of these two types. Consider first a
diagram with $d$ MHV tree vertices. Each vertex has two negative
helicity gluons. To make a connected tree diagram, we must connect
the vertices with $d-1$ propagators, absorbing $d-1$ negative
helicity gluons (as each propagator absorbs a negative helicity
gluon at one end). This leaves $d+1$ negative helicity external
gluons. To make a diagram with $l$ loops, we need $l$ additional
propagators, leaving \eqn\nonn{q=d+1-l} external gluons of
negative helicity. This formula, which has already been explained
in \CachazoKJ, is illustrated in \figtor(a). While keeping the
degree $d$ fixed, we can replace an MHV tree vertex with a
one-loop all-plus vertex; to also keep the number $l$ of loops
fixed, we should remove one propagator. See \figtor(b) for an
example of such a diagram. In this process, the change in the
vertex reduces $q$ by 2, while removing a propagator increases it
by 1.  So overall, $q$ is reduced by 1 and \nonn\ becomes
\eqn\unonn{q=d+1-l-p,} where $p$ is the number of all-plus
vertices. The possible range of $p$ is $0\leq p\leq l$, so we can
also write an inequality \eqn\degi{q-1+l\leq d\leq q-1+2l.}

\subsec{The $-++\dots ++$ One Loop Amplitude}

Now we will compare this guess to the actual behavior of the
nonsupersymmetric amplitudes with a single gluon of negative
helicity.  This comparison has revealed both good news and bad
news. The good news is that the twistor space structure agrees
with what we would expect.  The bad news is that we have not been
able to find an off-shell continuation of the all-plus scattering
amplitude to give the right amplitude.

The one-loop nonsupersymmetric scattering amplitudes in which all
gluons  but one have the  same helicity have been derived  by
Mahlon \refs{\mahlona, \mahlonb} using recursive techniques. For
example, the five-gluon one-loop amplitude with helicities $-++++$
is \refs{\BernSX,\bernfive} \eqn\fiveppppm{A={i\over 48
\pi^2}{1\over \vev{34}^2}\left[-{[25]^3 \over
[12][51]}+{\vev{14}^3[45]\vev{35} \over \vev{12} \vev{23}
\vev{45}^2}-{\vev{13}^3[32]\vev{42} \over \vev{15} \vev{54}
\vev{32}^2} \right].}  The product of any three coplanar operators
annihilates this amplitude \eqn\rrr{K^3A=0.} This can be proved as
follows. Using the homogeneity of $K$ in the
$\partial/\partial\tilde\lambda$'s, it follows that $K^2A$ is
homogeneous of degree $-1$ in inner products $[i,j]$ of negative
helicity spinors.  In section 3 of \WittenNN, it is shown that any
momentum-conserving five-gluon amplitude, such as $K^2A$, that is
homogeneous of degree $-1$ in the $[i,j]$, is annihilated by $K$.
So $K^3A=0$. We have also verified with computer assistance that
$A$ is annihilated by a certain product of collinear operators,
\eqn\pahi{F_{234}^3F_{345}^3A=0.}

\ifig\gip{Two representations of a diagram contributing to a
one-loop nonsupersymmetric amplitude with helicities $-++++$. (a)
The twistor-space geometry as found from the differential
equations. The two lines intersect as shown.  (b) A hypothetical
representation of the amplitude in terms of a tree diagram with a
four-valent all-plus vertex and a three-valent MHV tree vertex. }
{\epsfxsize=0.75\hsize\epsfbox{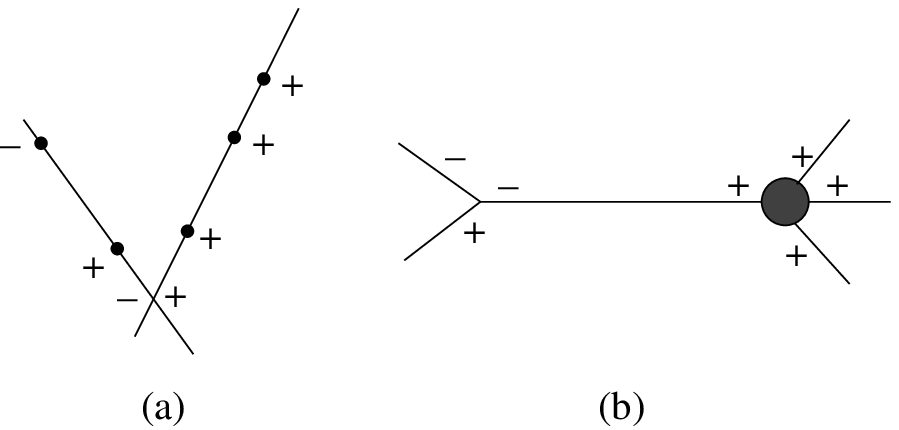}}

These statements correspond  to the twistor space picture of
\gip(a).  (The picture must be infinitesimally thickened to allow
for ``derivative of a delta function support,'' because of the
cubes in \rrr\ and \pahi.)  All five gluons are contained in a
plane because of \rrr. In addition, \pahi\ asserts that three of
them (2, 3, and 4 or 3, 4, and 5 -- in other words, three
consecutive positive helicity gluons) are contained in a line.
Drawing a straight line through the other two points, we find that
the five gluons are on a union of two intersecting lines, as
indicated in the figure. This figure is in agreement with what we
would expect from the Feynman diagram of \gip(b), with an MHV
vertex and an all-plus vertex (and a $1/P^2$ propagator, whose
principal part describes an intersection of the two lines in
twistor space, as we explained at the end of section 2).  One
subtlety is worth noting. In \gip, all contributions have external
helicities $+++$ on one line (or vertex) and $+-$ on the other;
there is no contribution with $++$ on one side and $++-$ on the
other side. This is what we would expect if the all-plus vertices
are $n$-valent with $n\geq 4$, in other words if the all-plus
vertex with $n=3$ vanishes off-shell just as it does on-shell.

We have similarly studied the $-+++\dots ++$ amplitudes with up to
eight gluons and found that they obey differential equations that
are consistent with the twistor space picture of \gip(a) (with the
extra gluons added on one line or the other, preserving the cyclic
order).

\bigskip\noindent{\it Off-Shell Continuation Of The All-Plus
Amplitude?}

On the other hand, we have not been successful in finding an
off-shell continuation of the all-plus one-loop amplitudes to use
in diagrams such as \gip(b). To find this off-shell continuation,
one approach we considered was to take the second version
\allplusb\ of the all-plus amplitude. This only depends on the
momenta of external lines, which are defined off-shell, and on the
positive chirality spinors $\lambda$, which we continued off-shell
in \CachazoKJ. So this gives a candidate off-shell continuation of
the all-plus amplitude, but we have found that when inserted in
Feynman diagrams such as that of \gip(b), it does not lead to the
right scattering amplitudes. A similar problem in defining
appropriate off-shell continuations for MHV gravity amplitudes was
found in \refs{\GiombiIX,\WuFB}.

An interesting point here is that given the absence of a one-loop
$+++$ vertex, it is impossible to write a tree diagram with MHV
and one-loop all-plus vertices that contribute to the one-loop
$-+++$ amplitude.  Yet the one-loop $-+++$ amplitude is nonzero.
It therefore is necessary, from this point of view, to interpret
the one-loop $-+++$ amplitude as a new local vertex.  Indeed, it
can be shown that this amplitude is annihilated by $F^2$ and so is
supported on a line in twistor space; it is thus at least somewhat
natural to interpret it as a new vertex.

\subsec{Nonsupersymmetric ``MHV'' Amplitudes}

Here we consider the non-supersymmetric one-loop amplitudes
 with two gluons of negative helicity.  We might call them
``MHV'' amplitudes, but in the nonsupersymmetric case this name
does not fit well, since the amplitudes with less than two
negative helicity gluons are also nonzero, as we have just
reviewed.

These $--++\dots +$ amplitudes contain cuts that can be determined
from unitarity.  According to \BernCG, the cut-constructible part
of the scalar one-loop amplitude with gluons 1 and 2 having
negative helicity (and the others positive helicity) is
\eqn\mhvsc{\eqalign{A_{sc,\, {\rm cut}}&={1\over 3} A^{{\cal N}=1}
-{c_\Gamma \over 3}A^{\rm tree}\sum_{\p=4}^{n-1}
{L_2\left(t_2^{[\p-2]}/t_2^{[\p-1]}\right) \over \left(t_1^{[2]}
t_2^{[\p-1]}\right)^3} \cr &\times\tr_+[\slashed{p}_1
\slashed{p}_2 \slashed{p}_\p \slashed{q}_{\p,1}]
\,\tr_+[\slashed{p}_1 \slashed{p}_2 \slashed{q}_{\p,1}
\slashed{p}_\p]\,\bigl(\tr_+[\slashed{p}_1 \slashed{p}_2
\slashed{p}_\p \slashed{q}_{\p,1}]-\tr_+[\slashed{p}_1
\slashed{p}_2 \slashed{q}_{\p,1} \slashed{p}_\p]\bigr),}} where
\eqn\ltwo{L_2(x)={\ln(x)-(x-1/x)/2 \over (1-x)^3}.}

Setting $P=p_{\p+1}+p_{\p+2}+\dots+p_1$ and
$Q=p_2+p_3+\dots+p_{\p-1},$ and writing $A^{\rm tree}$ for the
tree level MHV amplitude, the scalar one-loop amplitude becomes
\eqn\mhvsi{\eqalign{A_{sc, \, {\rm cut}}&= {1 \over 3} A^{{\cal
N}=1}- {c_{\Gamma} \over 3} {A^{\rm tree} \over \vev{1,2}^3}
\sum_{\p=4}^{n-1} {L_2(P^2/Q^2) \over (Q^2)^3} \cr & \times
\vev{1,\p}\vev{2,\p} \langle1|P|\p]\langle2|P|\p]\bigl(\vev{1,\p}
\langle2| P|\p] -\vev{2,\p}\langle1|P|\p]\bigr). \cr }} (As in
section 3, we write $\langle\lambda|P|\mu]=\lambda^aP_{a\dot
a}\mu^{\dot a}$.) Now we have two different triangle functions,
\eqn\trif{T(p,P,Q)={\ln(Q^2/P^2) \over Q^2-P^2}, \qquad \tilde
T(p,P,Q) = {L_2(P^2/Q^2) \over (Q^2)^3}.}

Schematically, the amplitude is a sum of triangle diagrams
\eqn\amt{A_{sc,\, {\rm cut}}={1\over 3}\sum_{\p = 4}^{n-1}
c^{1,2}_{1,\p} T(p_\p,P,Q)+\sum_{\p =4}^{n-1}\tilde{c}^{1,2}_{\p}
\tilde{T}(p_\p,P,Q),} where the first sum gives ${1\over
3}A^{{\cal N}=1}$ and $\tilde c_{\p}^{1,2}$ is the coefficient in
front of $\tilde T(p_\p,P,Q)$ in \mhvsi.

\bigskip
\ifig\zeromm{A triangle diagram contributing to the scalar loop
amplitude with adjacent negative helicity gluons.
}{\epsfxsize=0.50\hsize\epsfbox{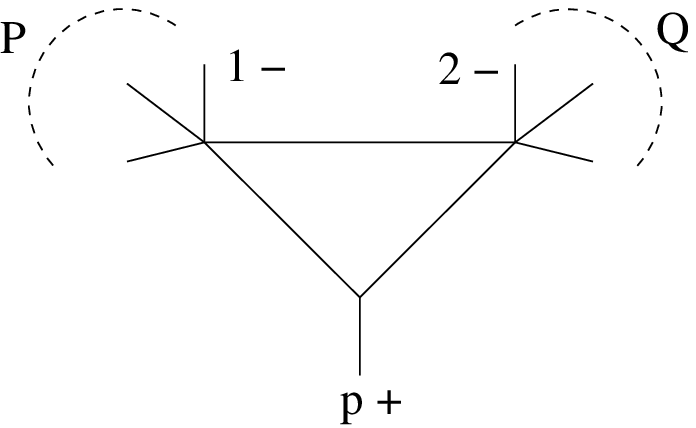}}

The first part of the amplitude, involving $T(p,P,Q)$, was studied
in section 4. The part of the amplitude involving the
nonsupersymmetric triangle function $\tilde T(p,P,Q)$ has almost
the same twistor space structure. The gluons whose momenta add to
$P$ are contained in a line $L$  in view of the criterion of
section 2.1, and the gluons whose momenta add to $Q$ are likewise
contained in a line $L'$. The $\tilde T$ part of the amplitude is
annihilated by $K^2$ (as we have found with some computer
assistance), so all gluons are coplanar and in particular $L$ and
$L'$ intersect. Since the amplitude is annihilated by $K^2$ but
not by $K$, it has ``derivative of a delta function'' support for
such coplanar configurations; the nature of this ``derivative of a
delta function'' support is further constrained by an additional
equation that is analogous to \sop: \eqn\nofhog{
K_{PPQQ}F^2_{pPP}F^2_{pQQ}\left(\tilde c^{i,j}_{\p,a}\tilde
T(p,P,Q)\right)=0.}

\bigskip\noindent{\it Cut-Free Terms}

These cut-constructible terms do not give the full
non-supersymmetric $--+++\dots +$ amplitude. In particular, they
lack poles in certain multiparticle channels. The missing parts of
the amplitudes are cut-free rational functions. For five gluons
with helicities $--+++$, the rational function has been computed
via string-inspired methods \bernfive.\

With computer assistance, we have found that this rational
function is annihilated by $K^2$, and so corresponds in twistor
space to a planar configuration.  Moreover, the rational function
is annihilated by $F_{234}^2F_{345}^2F_{451}^2$, and so is
supported on configurations on which three gluons,  including at
most one of negative helicity, are collinear. It is not
annihilated by $K_{1235}F_{234}^2F_{145}^2$, which (according to
\nofhog) annihilates the cut-constructible part of the amplitude.
These two facts can be understood if we assume again that there is
no one-loop $+++$ vertex and that there is a one-loop $-++ +$
vertex.

\ifig\posi{All possible diagrams contributing to the cut-free part
of the full non-supersymmetric $--+++$ amplitude.}
{\epsfxsize=0.75\hsize\epsfbox{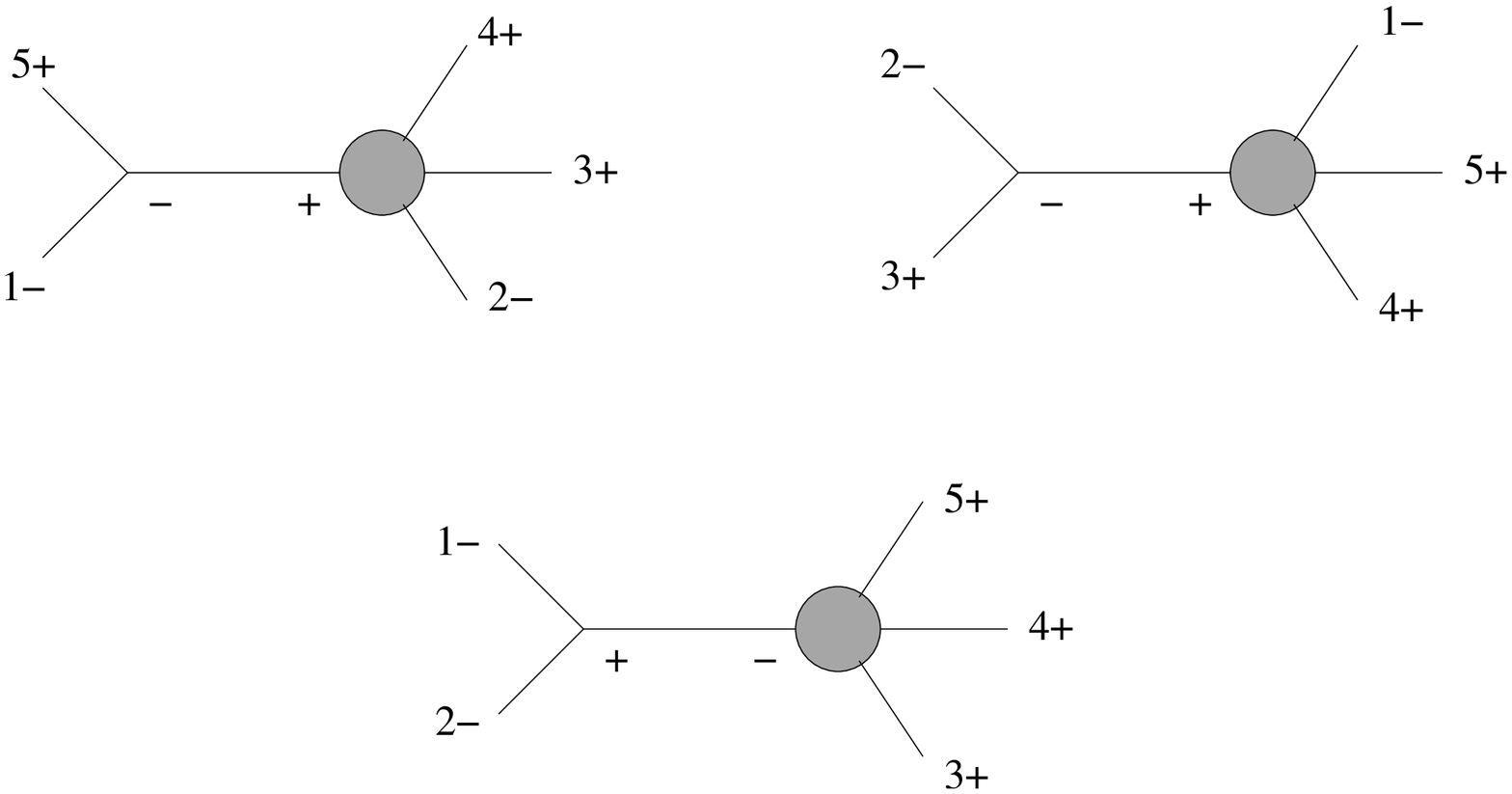}}

According to this, the cut-free part of the $--+++$ amplitude
receives contributions from only three configurations, shown in
\posi. Each diagram has two vertices: one trivalent MHV vertex and
one $-+++$ vertex. By definition, the trivalent MHV vertex must
contain at least one external gluon of negative helicity and
therefore the $-+++$ vertex has at most one external gluon of
negative helicity. The collinear operators are squared because the
$-+++$ vertex is localized on a line with a derivative of a delta
function support. This also explains why
$K_{1235}F_{234}^2F_{145}^2$ does not annihilate the amplitude.
The collinear operators $F_{234}^2$ and $F_{145}^2$ annihilate the
two configurations on the top in \posi, leaving the one with
gluons $3,4,5$ on the $-+++$ vertex. Due to the derivative of a
delta function support, a single $K_{1235}$ is not enough to
annihilate the diagram and it should be supplemented by an extra
collinear operator $F_{345}$.

\ifig\genp{ A diagram that could contribute to the rational
function part of the $--+++$ loop amplitude if a $+++$ one-loop
vertex is included.}
{\epsfxsize=0.55\hsize\epsfbox{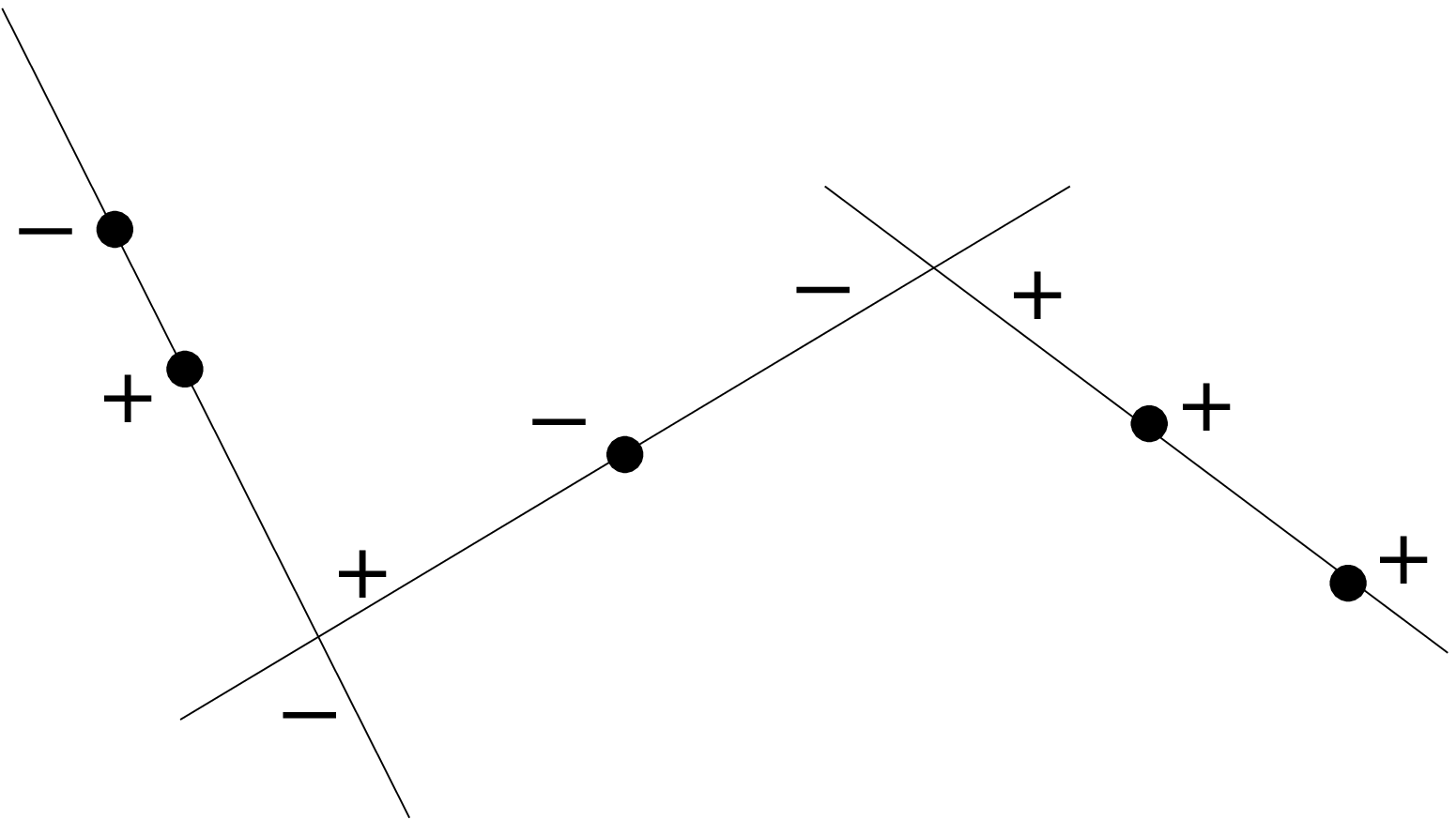}}

The fact that the rational function part of the $--+++$ amplitude
is annihilated by $K^2$ gives further evidence that there is no
one-loop $+++$ vertex, even off-shell. If such a vertex exists,
the one-loop nonsupersymmetric amplitude $--+++$ can receive a
contribution from a tree diagram with two MHV vertices and one
all-plus vertex (\genp).   This configuration is not planar, so
its contribution would not be annihilated by $K^2$.

 Unfortunately,
the cut-free parts of non-supersymmetric one-loop ``MHV''
amplitudes have not yet been computed for $n>5$ gluons. These
amplitudes should receive contributions from the quiver of \genp\
(with additional positive helicity gluons placed on the all-plus
line).  So their cut-free part should not be annihilated by $K^2$,
but should obey differential equations reflecting the structure of
this quiver.

\bigskip
\bigskip
\centerline{\bf Acknowledgements}

It is a pleasure to thank Z. Bern, L. Dixon, A. De Freitas and V.
Voevodsky for helpful discussions. Work of F. Cachazo was
supported in part by the Martin A. and Helen Chooljian Membership
at the Institute for Advanced Study and by DOE grant
DE-FG02-90ER40542; that of P. Svrcek by NSF grants PHY-9802484 and
PHY-0243680; and that of E. Witten by NSF grant PHY-0070928.
Opinions and conclusions expressed here are those of the authors
and do not necessarily reflect the views of funding agencies.

\appendix{A}{Box Functions}

The box function $F_{n:r;i}^{\rm 2m~e}$ is one of a set of
functions constructed from the scalar box integrals. The latter
form a complete list of the possible integrals that can appear in
a Feynman diagrammatic computation of one-loop amplitudes in ${\cal
N}=4$ gauge theory.\foot{After Passarino-Veltman reduction formulas
are applied.}

These integrals are known as the scalar box integrals because they
would arise in a one-loop computation of a scalar field theory
with four internal propagators.

\ifig\pipo{Scalar Box Integrals used in the definition of: (a) The
box function $F_{n:r;i}^{\rm 2m~e}$. (b) The generic box function
$F(p,q,P)$.} {\epsfxsize=0.85\hsize\epsfbox{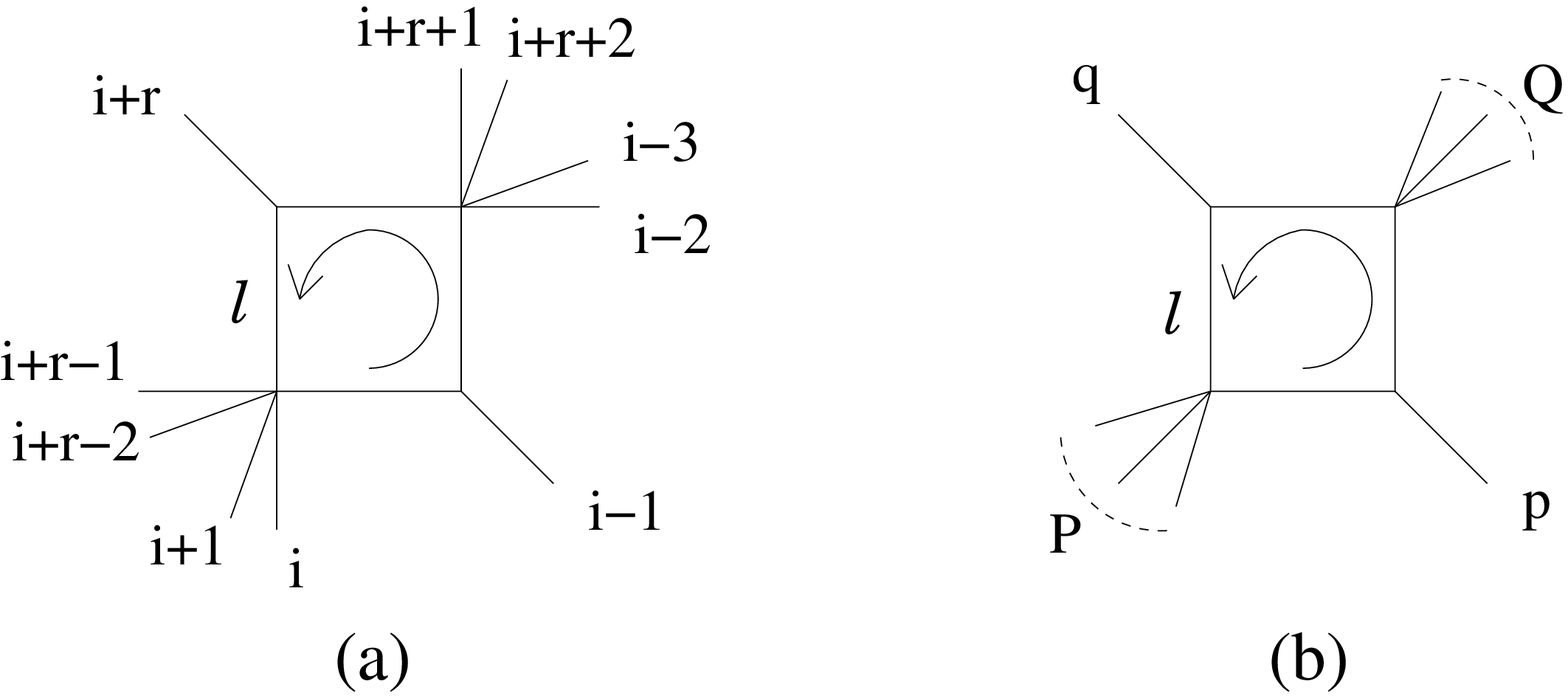}}

The scalar box integral is defined as follows:
\eqn\bbo{ I_4 = -i(4\pi)^{2-\epsilon} \int
{d^{4-2\epsilon}\ell\over (2\pi)^{4-2\epsilon}} {1\over \ell^2
(\ell-K_1)^2(\ell-K_1-K_2)^2(\ell+K_4)^2}. }
The incoming external momenta at each of the vertices are
$K_1,K_2,K_3,K_4$. The labels are given in consecutive order
following the loop. Momentum conservation implies that
$K_1+K_2+K_3+K_4 = 0$ and this is why \bbo\ only depends on three
momenta.

We are interested in the case when $K_1 = p_{i-1}$, $K_2 =
p_i+\ldots +p_{i+r-1}$ and $K_3=p_{i+r}$ (see figure 1A).

The box function \scalar\ is then defined as follows,
\eqn\sbox{ F_{n:r;i}^{2m e} = \left( t_{i-1}^{[r+1]} t_i^{[r+1]} -
t_i^{[r]}t_{i+r+1}^{[n-r-2]} \right) I_{4:r;i}^{2m e} }
Here we follow the notation in \BernZX, which is motivated by the
fact that $K_2$ and $K_4$ are not lightlike and can be thought of
as  momenta of massive scalar particles. $({\rm 2m~e})$ stands for
``two masses" and ``easy." The ``easy" case is when the two masses
are at diagonally opposed corners. The ``hard" case $({\rm 2m~h})$
is when the masses are adjacent. Fortunately, the latter does not
enter in one-loop MHV amplitudes but it does for the six-gluon
non-MHV one-loop amplitudes \BernCG.

In section 3.1, we introduced the generic box function $F(p,q,P)$.
This can be defined in a similar way by using the assignment of
momenta shown in figure 1B.

\appendix{B}{Proof of Coplanarity of Lines in $H_q(p,P)$}

The main result of section 3.2 was the decomposition of the box
function \scalk,
\eqn\demm{\eqalign{  & F(p,q,P)  = -{1\over \epsilon^2}\left[
(-(P+p)^2)^{-\epsilon} + (-(P+q)^2)^{-\epsilon}
-(-P^2)^{-\epsilon} -(-Q^2)^{-\epsilon}\right] \cr & + H_q(p,P) +
H_p(q,P) + H_q(p,Q) + H_p(q,Q)  - \log\left( {\vev{p,Q}\over
\vev{p,P}}\right) \log\left( {\vev{q,P} \over \vev{q,Q}} \right)
}}
where
\eqn\lapp{\eqalign{ H_q (p,P) =&  - {\rm Li}_2 \left( 1 -
{\vev{p,P+q}\over \vev{p,P}}{P^2\over (P+q)^2} \right) + {\rm
Li}_2 \left( 1 - {P^2\over (P+q)^2} \right) \cr & + \log \left(
{\vev{p,P+q}\over \vev{p,P}} \right) \log (P+q)^2 -{1\over
4}\log^2\left( {\vev{p,P+q}\over \vev{p,P}} \right).}}

The twistor transform of the function $H_q(p,P)$ was shown to be
localized on a configuration where gluons in $\tilde P$ are on a line
$L$, and gluons in $\tilde Q$ are
on another line $L'$. It was further  claimed that the two lines
intersect and moreover that the remaining gluon with momentum $q$ is
contained (with derivative of delta function support)
in the plane defined by the two lines.

In this appendix, we provide a proof of the fact that the two lines
intersect. More precisely, we prove the equivalent statement that
the points corresponding to two gluons in $\tilde P$ and two gluons in
$\tilde Q$ are coplanar. That is, we prove that
\eqn\state{K_{P_1,P_2,Q_1,Q_2} H_q(p,Q) = 0.}
where $K_{ijkl}$ is the differential operator of degree two
obtained from the geometric condition of coplanarity \urov.
$(P_1,P_2)$ are any to gluons in $\tilde P$ and $(Q_1,Q_2)$ are any two
gluons in $\tilde Q$. Using conformal invariance, we can set the twistor
space coordinates of $Q_1$ and $Q_2$ to be $Z_{Q_1} =(1,0,0,0)$
and $Z_{Q_2} = (0,1,0,0)$. This reduces $K_{P_1,P_2,Q_1,Q_2}$ to
\eqn\newK{ K = \epsilon^{\dot a\dot b}{\del^2\over \del
\tilde\lambda_{P_1}^{\dot a}\del \tilde\lambda_{P_2}^{\dot b}} }

Before getting into the proof of \state, let us note some useful
facts about the  dilogarithm and its derivatives. Using the definition
of the dilogarithm, ${\rm Li}_2(x) = -\int^x_0 \log (1-z)dz/z$, it
is easy to see that
\eqn\deri{ {d\over dx}{\rm Li}_2\left(1-x\right) = {\log x\over
1-x}.}
Moreover, upon using the chain rule one can show that if
$X=X(y,z)$, then any differential operator ${\cal O}$ that is
homogeneous and degree two in $\partial/\partial y$ and
$\partial/\partial z$  acts to produce,
\eqn\hodi{ {\cal O}\!\left( {\rm Li}_2\left(1-X\right) \right) =
-\log X ~ {\cal O}\!\left(\log (1-X)\right) + \ldots }
where $\ldots$ denotes rational functions of $X$ and its
derivatives.

The proof of \state\ will proceed in three steps. First, we prove
that the terms  in $K H_q(p,P)$ containing logarithms vanish.
Second, we prove that the rational part of $K {\rm Li}_2 \left( 1
- {P^2\over (P+q)^2} \right)$ is zero. And third, we prove that
the rational functions from the remaining two terms in \lapp\
cancel each other.

\subsec{Logarithmic Terms}

Using \hodi\ and \lapp, we find that
\eqn\lote{ \eqalign{ & K \left[ H_q(p,P)\right]_{\rm log} =
\log\left( {\vev{p,P+q}\over \vev{p,P}}{P^2\over (P+q)^2} \right)
K \left[\log\left( 1-{\vev{p,P+q}\over \vev{p,P}}{P^2\over
(P+q)^2} \right)\right] \cr & - \log\left( {P^2\over (P+q)^2}
\right)K \left[\log\left( 1-{P^2\over (P+q)^2} \right)\right]
+\log \left( {\vev{p,P+q}\over \vev{p,P}} \right) K \left[ \log
(P+q)^2\right] . }}
We have not included the term where $K$ acts on $\log \left(
{\vev{p,P+q}\over \vev{p,P}} \right)$, since it vanishes. To see
this note that $\eta$ can be chosen to be $\eta^{\dot
a}=\delta^{\dot a \dot 2}$ or $\eta^{\dot a}=\delta^{\dot a \dot
1}$ so that only one and the same component of
$\tilde\lambda_{P_1}$ and $\tilde\lambda_{P_2}$ appears. But $K$,
given in \newK, only contains mixed terms.

The only way $K \left[ H_q(p,P)\right]_{\rm log}$ can vanish is if
the coefficient of each of the independent logarithms vanishes.
{}From \lote\ we see that there are only two independent
logarithms, namely, $\log\left( {\vev{p,P+q}\over
\vev{p,P}}\right)$ and $\log\left( {P^2\over (P+q)^2} \right)$.

The coefficient that multiplies the first is
\eqn\cofi{ K \left[\log\left( 1-{\vev{p,P+q}\over
\vev{p,P}}{P^2\over (P+q)^2} \right)\right] + K \left[ \log
(P+q)^2\right], }
while the coefficient of the second is
\eqn\cose{ K \left[\log\left( 1-{\vev{p,P+q}\over
\vev{p,P}}{P^2\over (P+q)^2} \right)\right] - K \left[\log\left(
1-{P^2\over (P+q)^2} \right)\right]. }

The task at hand is to prove that \cofi\ and \cose\ are zero.
However, note that \cose\ can be written as \cofi\ minus
\eqn\newco{ K\left[ \log\left( (P+q)^2 - P^2\right) \right].}
Therefore, after proving the vanishing of \cofi\ we are left with
proving that \newco\ is zero.

As a warm up, let us prove first that \newco\ is zero. The argument of the logarithm
equals $2 P\cdot q$. Let us write
$P = P_1+P_2+\hat P$. Note that $2 P\cdot q =
\vev{P_1,q}[P_1,q]+\vev{P_2,q}[P_2,q]+ 2\hat P\cdot q$ is linear
in $\tilde\lambda_{P_1}^{\dot a}$ and $\tilde\lambda_{P_2}^{\dot
a}$. This implies that $K (2P\cdot q) = 0$. Therefore,
\eqn\qino{ K\left[ \log\left( 2P\cdot q \right) \right] = -{1\over
(2P\cdot q)^2}\epsilon^{\dot a\dot b}{\del \over
\del\tilde\lambda_{P_1}^{\dot a}}(2P\cdot q) {\del \over
\del\tilde\lambda_{P_2}^{\dot b}}(2P\cdot q) =
-{\vev{P_1,q}\vev{P_2,q}\over (2P\cdot q)^2}[q,q] = 0.}

In order to prove that \cofi\ vanishes, we first add to it zero in
the form $K \left[ \log\left( {\vev{q,P}\over \vev{q,P}} \right)
\right]$. Then, we combine the terms as follows
\eqn\combi{ K\left[ \log\left(
{\vev{p,P}(P+q)^2-\vev{p,P+q}P^2\over \vev{q,P}} \right) \right] +
K\left[ \log\left({\vev{q,P} \over \vev{p,P} }\right) \right].}
We have encountered the argument of the first logarithms before; it
is equal to $(s\alpha + t\beta)/(u'\alpha + v'\beta)$, which was
shown to be independent of $\eta$ in the proof of \factor.
Therefore we can choose $\eta = \tilde\lambda_q$ to evaluate it.
This gives $\left\langle p | P | q\right]$. On the other hand, the
second term in \combi\ is trivially zero for the same reason that
$K\left[ \log\left({\vev{p,P+q} \over \vev{p,P} }\right) \right]$
vanishes, as discussed above.

So we are left with proving that
\eqn\jiko{ K\left( \log \left\langle p | P | q\right] \right) = 0
.}
A computation similar to \qino\ reveals that this is also
proportional to $[q,q] =0$.

This concludes the proof of the vanishing of \lote.

\subsec{Rational Terms}

As discussed before, this part of the proof is divided into two
computations. First, we  prove that
\eqn\rato{ K \left[ {\rm Li}_2 \left( 1 - {P^2\over (P+q)^2}
\right) \right]_{\rm rational} = 0,  }
where the subscript means the rational part. Sometimes we will
simply write ``r" instead of the whole ``rational" subscript.

Clearly, the rational part is obtained when one derivative acts to
produce a logarithm times a rational function and the second
derivative acts on the logarithm to produce a rational function.
More explicitly, we have
\eqn\expi{K \left[ {\rm Li}_2 \left( 1 - {P^2\over (P+q)^2}
\right) \right]_{\rm rational} \sim ~ \epsilon^{\dot a\dot b}{\del
\over \del\tilde\lambda_{P_1}^{\dot a}}\left({P^2\over
(P+q)^2}\right) {\del \over \del\tilde\lambda_{P_2}^{\dot
b}}\left({P^2\over (P+q)^2}\right)  }
where $\sim$ means equal up to an irrelevant rational function. By
writing ${P^2\over (P+q)^2}$ as  $1 - {2P\cdot q\over  (P+q)^2}$
we find, after a straightforward computation similar to \qino,
that
\eqn\sopu{ \epsilon^{\dot a\dot b}{\del \over
\del\tilde\lambda_{P_1}^{\dot a}}\left({2P\cdot q \over
(P+q)^2}\right) {\del \over \del\tilde\lambda_{P_2}^{\dot
b}}\left({2P\cdot q\over (P+q)^2}\right) = 0.}
This concludes the proof of \rato.

Finally, we have to prove that
\eqn\fino{ K\left[ -{\rm Li}_2 \left( 1 - {\vev{p,P+q}\over
\vev{p,P}}{P^2\over (P+q)^2} \right) + \log \left(
{\vev{p,P+q}\over \vev{p,P}} \right) \log (P+q)^2 \right]_{\rm
rational} = 0.}

In principle, a direct computation of each term should provide a
proof of \fino. However, it turns out that the dilogarithm in
\fino\ leads to a large proliferation of terms. Therefore, we seek
an alternative way of computing it.

Let us start by using Landen's identity \idone\ on both
dilogarithms of $H_q(p,P)$ to get
\eqn\newH{ \eqalign{ H_q & (p,P) = {\rm Li}_2 \left( 1 -
{\vev{p,P}\over \vev{p,P+q}}{(P+q)^2\over P^2} \right) - {\rm
Li}_2 \left( 1 - {(P+q)^2\over P^2} \right) \cr & - \log \left(
{\vev{p,P}\over \vev{p,P+q}} \right) \log P^2 + {1\over
4}\log^2\left( {\vev{p,P}\over \vev{p,P+q}}\right).}}

For any linear and homogeneous second order differential operator
${\cal O}$ and any function $X=X(y,z)$, the following is true:
\eqn\ticu{ {\cal O}\left[ {\rm Li}_2 \left( 1-{1\over X} \right)
\right]_{\rm rational} = -{1\over X}\; {\cal O}\left[ {\rm Li}_2
\left( 1-X \right)\right]_{\rm rational}. }
This identity together with \rato\ implies that
\eqn\good{ K \left[ {\rm Li}_2 \left( 1 - {(P+q)^2\over P^2}
\right) \right]_{\rm rational} = 0.}
Therefore, applying $K$ to both representations of $H_q(p,P)$
given by \lapp\ and \newH\ and taking the rational part we
conclude that
\eqn\conse{ \eqalign{ & K\left[ {\rm Li}_2 \left( 1 -
{\vev{p~P}\over \vev{p~(P+q)}}{(P+q)^2\over P^2} \right) + \log
\left( {\vev{p~(P+q)}\over \vev{p~P}} \right) \log P^2
\right]_{\rm rational} = \cr &  K\left[ -{\rm Li}_2 \left( 1 -
{\vev{p,P+q}\over \vev{p,P}}{P^2\over (P+q)^2} \right) + \log
\left( {\vev{p,P+q}\over \vev{p,P}} \right) \log (P+q)^2
\right]_{\rm rational} = 0}}
where we have used that $K\left[ \log^2\left( {\vev{p~P}\over
\vev{p~(P+q)}}\right) \right] = 0$.

Using the identity \ticu\ on the dilogarithm on the left we
produce a dilogarithm with the same argument as the one on the
right. Solving for it we find that
\eqn\compi{ K\left[ -{\rm Li}_2 \left( 1 - {\vev{p,P+q}\over
\vev{p,P}}{P^2\over (P+q)^2} \right) \right]_{\rm r} ={\alpha\over
(\alpha -1)} K\left[ \log \left( {\vev{p,P+q}\over \vev{p,P}}
\right) \log \left( {(P+q)^2\over P^2}\right) \right]_{\rm r}}
with
\eqn\defo{ \alpha = {\vev{p,P+q}\over \vev{p,P}}{P^2\over
(P+q)^2}. }

Using \compi\ to replace the complicated term with the dilogarithm
in \fino\ by a product of logarithms, we find the equivalent but
much simpler statement
\eqn\nova{\vev{p,P+q}P^2K\left[ \log \left( {\vev{p,P+q}\over
\vev{p,P}} \right) \log P^2 \right]_{\rm r} = \vev{p,P} (P+q)^2
K\left[ \log \left( {\vev{p,P+q}\over \vev{p,P}} \right) \log
(P+q)^2 \right]_{\rm r}.}
This new identity can be checked straightforwardly by explicit
computation.

\appendix{C}{Covariant Decomposition Of The Box Function.}

Our decomposition of the box function can be made manifestly
covariant by choosing $\eta^{\dot a}$ to be the
$\tilde\lambda^{\dot a}$ of one of the external gluons. Clearly,
different choices of $\eta^{\dot a}$ lead to very different
looking formulas. There are two particularly interesting choices
that reduce the total number of dilogarithms from eight to only
four. In general there are four $H$-functions and each contains
two dilogarithms. Choosing $\eta^{\dot a} = \tilde\lambda_p^{\dot
a}$ or $\eta^{\dot a} = \tilde\lambda_q^{\dot a}$ sets to zero two
of the $H$-functions. More explicitly we have:
\eqn\form{\eqalign{ \eta^{\dot a} = \tilde\lambda_p^{\dot a} ~
\Rightarrow ~ {\vev{q~(P+p)}\over \vev{q~P}} = {\vev{q~(Q+p)}\over
\vev{q~Q}} & = 1 ~ \Rightarrow ~ H_p(q,P) =H_p(q,Q) = 0, \cr
\eta^{\dot a} = \tilde\lambda_q^{\dot a} ~ \Rightarrow ~
{\vev{p~(P+q)}\over \vev{p~P}} = {\vev{p~(Q+q)}\over \vev{p~Q}} &
= 1 ~ \Rightarrow ~ H_q(p,P) =H_q(p,Q) = 0. } }

Consider the first choice, i.e., $\eta^{\dot a} =
\tilde\lambda_p^{\dot a}$. Then the box function \scalk\ is given by
\eqn\boxu{\eqalign{  F(p,q,P) & = -{1\over \epsilon^2}\left[
(-(P+p)^2)^{-\epsilon} + (-(P+q)^2)^{-\epsilon}
-(-P^2)^{-\epsilon} -(-Q^2)^{-\epsilon}\right] \cr &  - {\rm Li}_2
\left( {1+A {(1-B) \over (1-A)}} \right) - {\rm Li}_2 \left( {1+ B
{(1-A) \over (1-B)}} \right) - {1\over 2}\log^2\left( -{(1-A)\over
(1-B)}\right) \cr &  + {\rm Li}_2 \left( {1-C} \right)  + {\rm
Li}_2 \left( {1-D} \right)  + {1\over 2}\log^2\left( {C\over D}
\right)} }
where
\eqn\nota{ A = {P^2\over (P+p)^2}, ~ B ={Q^2\over (Q+p)^2}, ~ C
={P^2\over (P+q)^2}, ~ D ={Q^2\over (Q+q)^2}.}
Note that we have decreased the number of dilogarithm in the
original form of the box function \scalk\ by one.

\listrefs
\end